\documentclass[%
10pt,
reprint,
tightenlines,
superscriptaddress,
amsmath,
amssymb,
pra,
floatfix,
]{revtex4-2}
\date{\today}
\usepackage{subfiles}

\usepackage{fullpage}
\usepackage{etoolbox}
\apptocmd{\sloppy}{\hbadness 10000\relax}{}{}

\usepackage[english]{babel}
\usepackage{blindtext}

\usepackage{booktabs}
\usepackage{makecell}
\usepackage[table,xcdraw]{xcolor}
\usepackage{dcolumn}
\makeatletter
    \def\CT@@do@color{
      \global\let\CT@do@color\relax
            \@tempdima\wd\z@
            \advance\@tempdima\@tempdimb
            \advance\@tempdima\@tempdimc
    \advance\@tempdimb\tabcolsep
    \advance\@tempdimc\tabcolsep
    \advance\@tempdima2\tabcolsep
            \kern-\@tempdimb
            \leaders\vrule
                    \hskip\@tempdima\@plus  1fill
            \kern-\@tempdimc
            \hskip-\wd\z@ \@plus -1fill }
    \makeatother

\usepackage[caption=false, subrefformat=parens,labelformat=parens]{subfig}
\usepackage{graphicx}
\graphicspath{{Figures/}}
\newsavebox{\measurebox}
\usepackage{float}

\usepackage{ulem}
\usepackage{cancel}
\usepackage{siunitx}
\usepackage{amssymb}
\usepackage{amsmath}
\usepackage{verbatim}
\usepackage{soul}
\usepackage{bm}
\def\ket#1{| #1 \rangle}

\newcounter{prob_num}
\begin{document}
	\title{Isotope-Selective Laser Ablation Ion-Trap Loading of $\mathbf{^{137}\mathrm{Ba}^+}$ using a $\mathbf{\mathrm{BaCl}_2}$ Target}
	\author{Brendan M. White}
	\affiliation{Department of Physics and Astronomy, University of Waterloo, Waterloo, N2L 3R1, Canada}
	\affiliation{Institute for Quantum Computing, University of Waterloo, Waterloo, N2L 3R1, Canada}
	\author{Pei Jiang Low}
	\affiliation{Department of Physics and Astronomy, University of Waterloo, Waterloo, N2L 3R1, Canada}
	\affiliation{Institute for Quantum Computing, University of Waterloo, Waterloo, N2L 3R1, Canada}
	\author{Yvette de Sereville}
	\affiliation{Department of Physics and Astronomy, University of Waterloo, Waterloo, N2L 3R1, Canada}
	\affiliation{Institute for Quantum Computing, University of Waterloo, Waterloo, N2L 3R1, Canada}
	\author{Matthew L. Day}
	\affiliation{Department of Physics and Astronomy, University of Waterloo, Waterloo, N2L 3R1, Canada}
	\affiliation{Institute for Quantum Computing, University of Waterloo, Waterloo, N2L 3R1, Canada}
	\author{Noah Greenberg}
	\affiliation{Department of Physics and Astronomy, University of Waterloo, Waterloo, N2L 3R1, Canada}
	\affiliation{Institute for Quantum Computing, University of Waterloo, Waterloo, N2L 3R1, Canada}
	\author{Richard Rademacher}
	\affiliation{Department of Physics and Astronomy, University of Waterloo, Waterloo, N2L 3R1, Canada}
	\affiliation{Institute for Quantum Computing, University of Waterloo, Waterloo, N2L 3R1, Canada}
	\author{Crystal Senko}
	\email{csenko@uwaterloo.ca}
	\affiliation{Department of Physics and Astronomy, University of Waterloo, Waterloo, N2L 3R1, Canada}
	\affiliation{Institute for Quantum Computing, University of Waterloo, Waterloo, N2L 3R1, Canada}
	\begin{abstract}
		The $^{133}\mathrm{Ba}^+$ ion is a promising candidate as a high-fidelity qubit, and the $^{137}\mathrm{Ba}^+$ isotope is promising as a high-fidelity qudit ($d>2$). 
		Barium metal is very reactive, and $^{133}\mathrm{Ba}^+$ is radioactive and can only be sourced in small quantities, so the most commonly used loading method, oven heating, is less suited for barium, and is currently not possible for $^{133}\mathrm{Ba}^+$.
		Pulsed laser ablation solves both of these problems by utilizing compound barium sources, while also giving some distinct advantages, such as fast loading, less displaced material, and lower heat load near the ion trap. 
		Because of the relatively low abundances of the isotopes of interest, a two-step photoionization technique is used, which gives us the ability to selectively load isotopes. 
		Characterization of the ablation process for our $\mathrm{BaCl}_2$ targets are presented, including observation of neutral and ion ablation-fluence regimes, preparation/conditioning and lifetimes of ablation spots, and plume velocity distributions.
		We show that using laser ablation on $\mathrm{BaCl}_2$ salt targets with a two-step photoionization method, we can produce and trap barium ions reliably. 
		Further, we demonstrate that with our photoionization method, we can trap $^{137}\mathrm{Ba}^+$ with an enhanced selectivity compared to its natural abundance.
	\end{abstract}
	\maketitle
\section{Introduction}\label{sec:Intro}
    Ion traps have been studied extensively as a promising platform for quantum computing \cite{Duan2010, Bruzewicz2019_review}. 
    Trapped ions as qubits have been demonstrated with many different elements \cite{Ospelkaus2011, Dietrich2010, Ballance2016, Gaebler2016, Shapira2018, Webb2018, Christensen2020, Ballance2015, Tan2015, Bruzewicz2019_multi, Hughes2020}, and the barium isotopes, $^{133}\mathrm{Ba}^+$ and $^{137}\mathrm{Ba}^+$, show much promise for high-dimensional quantum information \cite{Hucul2017, Christensen2020, Low2020}. 
    However, barium presents some unique difficulties as an atomic source, because of its metal form's chemical properties, and $^{133}\mathrm{Ba}^+$ sourcing limitations. 
    To use these isotopes for quantum information, a reliable method must be developed for selectively trapping specific isotopes using barium-salt targets.
    
    Loading ions into an ion trap requires the generation of a flux of the desired atoms through the center of the trap, along with a method to ionize the atoms. 
    The most commonly used method for generating the atomic flux is Joule heating of an oven source \cite{Kjaergaard2000, Gulde2001, Devoe2002, Steele2007, Dietrich2010, Wang2011, Leschhorn2012, Graham2014, Ballance2018}. 
    
    In recent years, laser ablation has become a more common method for loading ions \cite{Vrijsen2019, Hucul2017, Olmschenk2017}.
    This method uses a pulsed laser with a high pulse energy to ablate atoms from a source \cite{Knight1981, Russo1995, Willmott2000, Ashfold2004, Phipps2007}, and it can result in less contamination \cite{Hendricks2007, Leibrandt2007}, a lower pressure rise at the trap center\cite{Hendricks2007}, and a lower overall heat load to the vacuum chamber \cite{Sheridan2011}.
    The major disadvantage of ablation is the high variance of neutral flux from shot-to-shot and for different spots on a target.
    Compared to ovens, which require metal sources (excepting special setups \cite{Devoe2002}), the source target in laser ablation can be either pure metal or a compound, and requires less material. 
    This amelioration solves the problem of generating $^{133}\mathrm{Ba}$ atoms, which can only be obtained in microgram quantities (due to its radioactive properties), and is typically obtained in the form of a chemical compound such as $\mathrm{BaCl}_2$. 
    Furthermore, ablation is useful for the element barium in general \cite{Schowalter2012, Hucul2017, Olmschenk2017}, which, in metal form, oxidizes within seconds in air \cite{Devoe2002, Graham2014}.
    For these reasons, we focus on laser ablation in our experiment.
    
    Using laser ablation, charged ions can be generated directly \cite{Knight1981, Hashimoto2006, Leibrandt2007, Davies2007, Zimmermann2012, Olmschenk2017, Hucul2017, Poschl2018, Sameed2020, Niemann2019, Kwong1989, Laska2003, Kwapien2007, Campbell2009}, or ablated neutral atoms can be ionized by electron-impact ionization \cite{Kornienko1999, Orient2002, Brown2007}. 
    Unfortunately, this method is significantly less efficient than other methods such as photoionization \cite{Gulde2001, Cetina2007}, and neither of these ionization methods discriminate between different isotopes.
    Resonant photoionization \cite{Gill1996, Yamada1988, Kjaergaard2000, Gulde2001, Lucas2004, Tanaka2005, Balzer2006, Steele2007, Brownnutt2007, Tanaka2007, Schuck2010, Sheridan2011, Leschhorn2012, Wang2011, Johanning2011, Graham2014, Guggemos2015, Zhang2017, Shao2018, Vrijsen2019}, in which multiple lasers are used to excite a valence electron and then eject it completely, is commonly used to selectively load ions. 
    Isotopic selectivity is important for applications such as quantum computing, which require the deterministic loading of 10 or more specific isotopes.
    This is a particularly salient point in the case of barium because of its large number of abundant, naturally-occurring isotopes.
    
    In this manuscript, we use laser ablation to generate neutral-barium atoms, and two-step photoionization to ionize them.
    We characterize ablation from a barium-salt target, describing the steps to prepare the target, and the best parameters we found for trapping ions consistently.
    A crucial step was ``conditioning'', where fresh spots on the target first needed to be ablated with a high fluence before they gave a sufficient atomic flux using a lower fluence.
    We present results of time-resolved spectroscopy on the resonant step of photoionization, giving an estimation of the velocities of ablated atoms. 
    We find that the ablation plume from a barium-salt target has a significantly higher temperature than that of metal targets. 
    Finally, we demonstrate trapping of $^{137}\mathrm{Ba}^+$, present loading rates, and show isotope selectivity for the ion of interest, $^{137}\mathrm{Ba}^+$. 
    We also discuss results from a low-volume ablation target intended for trapping $^{133}\mathrm{Ba}^+$, and we give a likely explanation for its insufficiency. 
\section{Experiment}\label{sec:Exp}
    \begin{figure*}
    	\centering
    	\includegraphics[width=\linewidth]{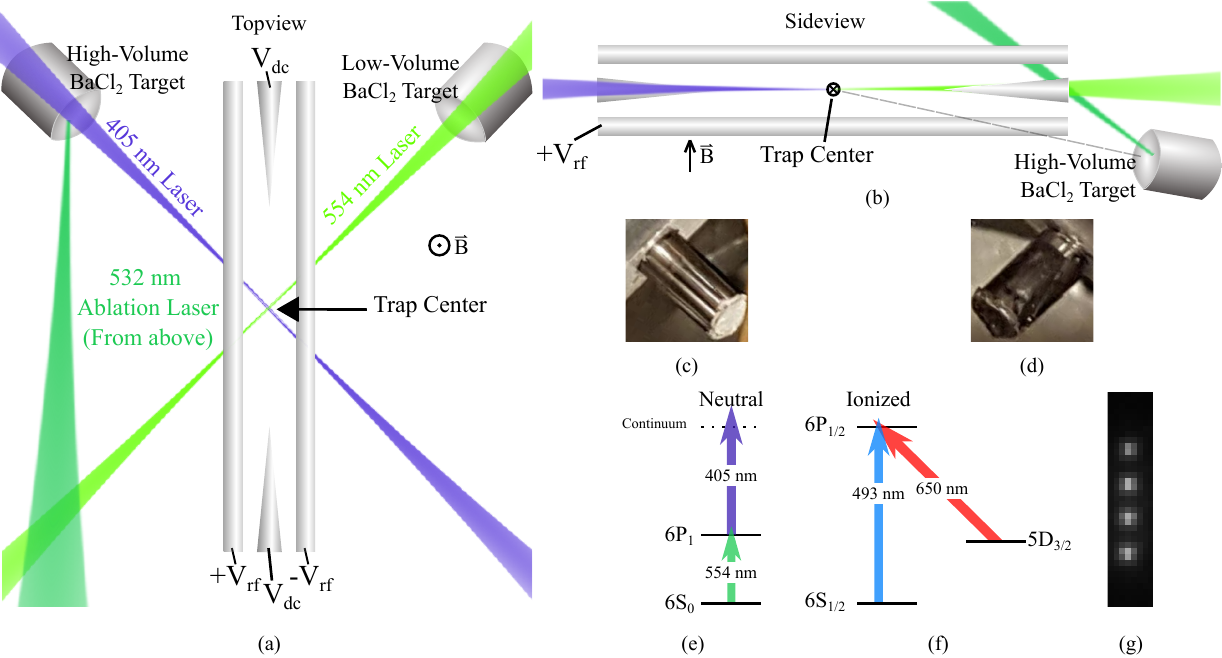}
    	\caption{(a-b) Experimental setup for laser ablation trapping. 
    		Ablation targets point directly towards the trap center. 
    		The ablation laser enters through the top viewport. 
    		To use the low-volume target, the ionization beams' directions are swapped. 
    		(c-d) High- and low-density targets respectively. 
    		(e) Relevant energy structure of neutral barium: a two-step photoionization process with $\SI{554}{\nano\meter}$ and $\SI{405}{\nano\meter}$ lasers \cite{Baird1979, Bekk1979, Niggli1987, Yamada1988, Wijngaarden1995, Kulaga2001, Leschhorn2012} is used to eject a valence electron from neutral barium, ionizing it.
    		(f) Relevant energy structure of ionized barium: a $\SI{493}{\nano\meter}$ laser drives the $S_{1/2}\leftrightarrow P_{1/2}$ transition to Doppler cool the ion, while a $\SI{650}{\nano\meter}$ repump laser drives the $D_{3/2}\leftrightarrow P_{1/2}$ transition. 
    		These cooling lasers are co-propogating with the first-step photoionization ($\SI{554}{\nano\meter}$) laser. (g) A chain of four trapped $^{138}\mathrm{Ba}^+$, with  $\sim\SI{13}{\micro\meter}$ distance between them.}
    	\label{fig:Ablation-Ionization-Setup}
    \end{figure*}
    The layout of our experiment is shown in Figure \ref{fig:Ablation-Ionization-Setup}(a-b). 
    The ion trap is a four-rod Paul trap with needles as end caps \cite{Low2019}. 
    Around \SIrange[range-units = single]{5}{10}{\volt} is applied to the needles to confine ions axially. 
    An rf voltage of $|V_{rf}| \approx \SI{66}{V}$ is applied to all four rods with each set of diagonal rods having opposite polarity, to confine ions radially. 
    
    The ablation targets are aluminum cylinders with recessed ends containing barium chloride ($\mathrm{BaCl}_2$), which are mounted inside the vacuum chamber below the side viewports.
    The targets' surfaces are angled directly towards the trap center in order to maximize the flux of trappable atoms. 
    The distance from the targets' surfaces to the trap center is $\SI{14.6}{\milli\meter}$. 
    The high-volume, natural-abundance target is used as a source for most isotopes of barium, while the low-volume target is used as a source for $^{133}\mathrm{Ba}$.
    
    Since barium metal oxidizes within seconds in air, we instead use barium salt. 
    For the high-volume target, we mixed natural-abundance $\mathrm{BaCl}_2$ with a small amount of deionized water to form a paste, which we applied to the recessed end of the ablation target, giving a high-volume, natural-abundance $\SI{30}{\milli\gram}$ $\mathrm{BaCl}_2$ target.
    The low-volume target was made from a solution of hydrochloric acid ($\mathrm{HCl}$), with approximately $\SI{0.4}{\milli Ci}$ of $^{133}\mathrm{Ba}$.
    Because $\mathrm{HCl}$ corrodes aluminum, we fit a tantalum foil over the tube's end. 
    To apply the solution to the target, we dropped $\SI{10}{\micro\liter}$ in the target cup (which is sitting on a $\SI{250}{\celsius}$ hotplate), let it evaporate, leaving $\SI{200}{\nano\gram}$ of $\mathrm{BaCl}_2$ salt on the surface, and repeated this process until we had $\sim\SI{10}{\micro\gram}$ on the substrate. 
    With such a low volume, the deposited salt is imperceptible on the substrate tantalum foil (see Figure \ref{fig:Ablation-Ionization-Setup}(c-d)).
    
    To ablate the targets, we use an Nd:YAG pulsed laser at wavelength $\SI{532}{\nano\meter}$ and pulse width \SIrange[range-units = single]{3}{5}{\nano\second}.
    We focus the ablation laser to a radius of $\SI{98}{\micro\meter}$ and $\SI{82}{\micro\meter}$ onto the high- and low-volume targets respectively. 
    With this beam size, we estimate that there are on the order of 500 distinguishable spots on the targets.
    The fluence is typically set from \SIrange[range-units = single]{0.2}{0.5}{\joule\per\cm\squared}. 
    In this paper, data is collected from ablation of the high-volume target unless otherwise noted. 
    The beam comes in from the top viewport at $\SI{56}{\degree}$ from the high- and low-volume target normals. 
    Two servomotors on a mirror mount in the ablation-laser path allow for precisely moving the ablation laser or sweeping over an area during an experiment.
    
    A two-step photoionization process, depicted in Figure \ref{fig:Ablation-Ionization-Setup}(e), is used to ionize ablated neutral atoms at the trap center. 
    The first step uses a $\SI{553.7}{\nano\meter}$ laser to drive one of the valence electrons of neutral barium into the $^1P_{1}$ state. 
    Since the first-step transition in neutral barium is resonant and has different frequencies for different isotopes, these peaks are discernible \cite{Baird1979, Bekk1979, Yamada1988} for a sufficiently low laser linewidth. 
    Therefore, in principle, this laser frequency can be varied to selectively ionize and load. 
    This laser is oriented perpendicular to the flux of neutral atoms in order to minimize Doppler broadening and shifting. 
    The second-step photoionization laser, at $\SI{405}{\nano\meter}$, is oriented perpendicular to the first-step laser to restrict the volume of ionization to be near the trap center, minimizing the potential energy of ions that are trapped and reducing the cooling time for hot, newly-captured ions. 
    The laser used to drive the first step of photoionization has a saturation power of $P_{sat} = \SI{0.57}{\mu W}$ \cite{White2019} (ionization lasers are focused to a $\SI{35}{\micro\meter}$ radius).
    
    Upon being ionized, ions begin to fluoresce from the $\SI{493}{\nano\meter}$ and $\SI{650}{\nano\meter}$ lasers, which are both co-propogating with the $\SI{554}{\nano\meter}$ laser. 
    The $\SI{493}{\nano\meter}$ laser is used to Doppler cool the ion, while the $\SI{650}{\nano\meter}$ laser is used to repump out of the $5D_{3/2}$ state.
    Once the ion has been cooled sufficiently, it crystallizes in the center of the ion trap. 
    Trapped ions are imaged with an imaging objective of $\mathrm{NA} = 0.26$. 
    
    The photons collected by the objective are read out using a PMT if the experiment requires measurement of the overall brightness of ions, or using a CMOS to spatially discriminate between ions in a crystallized chain. 
    The PMT can also be used to collect neutral fluorescence from ablated neutral-atom or ion flux.
\section{Ablation characterization}\label{sec:Ablation_Char}
    In this section, we describe preparation of fresh spots on the ablation target and the different regimes in which neutral atoms or ions are generated by ablation of our targets. 
    Further, we characterize spatial and temporal variation of atomic flux, and the typical number of pulses in which a single spot will give a significant atomic flux through the trap center. 
    
    We assess the atomic flux of neutral (or singly-ionized) barium atoms ablated through the trap center based on the fluorescence collected while the atoms transit through the $\SI{554}{\nano\meter}$ (or $\SI{493}{\nano\meter}$) laser beam, respectively. 
    The laser powers are set around 100-200x saturation and the laser frequencies are set to the resonance frequencies of the $^{138}\mathrm{Ba}^+$ isotope. 
    Fluorescence is quantified by the number of PMT counts in a  $\SI{55}{\micro\second}$ time window beginning $\SI{3}{\micro\second}$ after the ablation-laser pulse is incident on the ablation target. 
    As shown in Figure \ref{fig:Ablation-TOF-Spec}, this time window spans the majority of the observable neutral-fluorescence signal.
    Note that applying very high fluence, on the order of $\SI{1}{\joule\per\centi\meter^2}$ results in an appreciable amount of excited-state neutral atoms (or other fluorescing material) passing through the trap center, as discussed further in the Supplementary Materials. 
    For all results reported in this paper, we use laser fluences well below the threshold where this confounding signal appears.
    
    When aligned to a fresh spot on the target, no neutral-atom production is observed for fluences below $\SI{0.6}{\joule\per\centi\meter^2}$. 
    With a higher fluence of \SIrange[range-units = single]{0.6}{0.75}{\joule\per\centi\meter^2}, neutral-atom fluorescence of 100-400 counts starts to be observed. 
    Upon returning to low fluence, neutral-atom flux may then be observable. 
    We call this process of applying high-fluence pulses with the purpose of activating neutral atom production at low fluence, ``conditioning''. 
    New spots do not always give observable neutral fluorescence at high fluence, and even ones that do are not guaranteed to produce a significant neutral-fluorescence signal at low laser fluence, as shown later in this section.
    
    \begin{figure}
    	\centering
    	\includegraphics[width=0.99\linewidth]{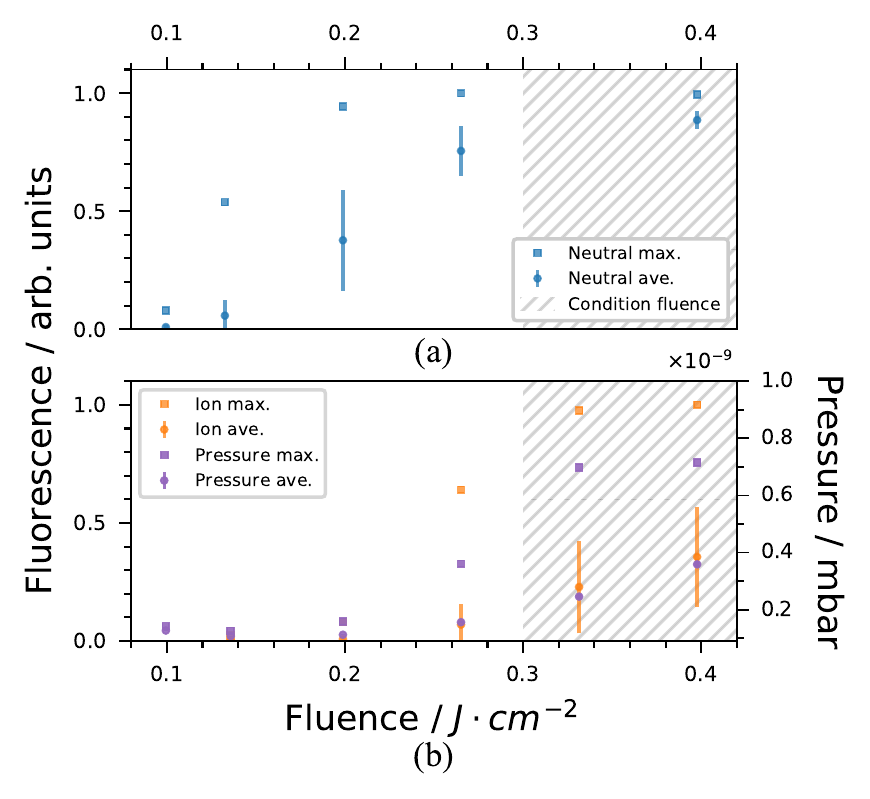}
    	\vspace{-15px}
    	\caption{Neutral atom and ion production regimes. Diagonally hatched areas denote conditioning fluences. (a) Neutral atom fluorescence vs pulse fluence. (b) Ion fluorescence and vacuum chamber pressure spikes vs pulse fluence.}
    	\label{fig:IonsvsNeutrals}
    \end{figure}
    Conditioning can allow access to a regime in which neutral atoms are produced but ionized atoms are not.
    As shown in Figure \ref{fig:IonsvsNeutrals}, neutral atoms are generated with a fluence above $\SI{0.1}{\joule\per\centi\meter^2}$, while ions are generated above $\SI{0.25}{\joule\per\centi\meter^2}$.
    The data in Figure \ref{fig:IonsvsNeutrals} was collected while sweeping the ablation laser over a $300\times\SI{700}{\micro\meter\squared}$ area which was previously conditioned with of order 100 pulses at $\SI{0.50}{\joule\per\centi\meter^2}$.
    Mean, maximum, and standard deviation are computed for data collected over two sweeps, with 120 pulses per sweep.
    We observe that the ion-fluorescence signal is well correlated spatially with the vacuum-chamber pressure spikes.
    The neutral fluorescence plateaus near the onset of ions, where there is much less spacial variation over the sweep than at lower fluence. 
    In practice, it is desirable to use spots that produce neutral atoms below the ions regime, to reduce excess charge build-up on electrodes and to more easily build chains of ions (see Section \ref{sec:Iso-Sel}).
    Because of the extreme variation during a sweep in this neutral-atom regime, the remaining data sets in this paper were collected with the ablation laser fixed on a single spot.
    
    \captionsetup[subfigure]{labelformat=empty}
    \begin{figure}
    	\centering
    	\subfloat[\label{sfig:Conditioning}]{\includegraphics[width=\linewidth]{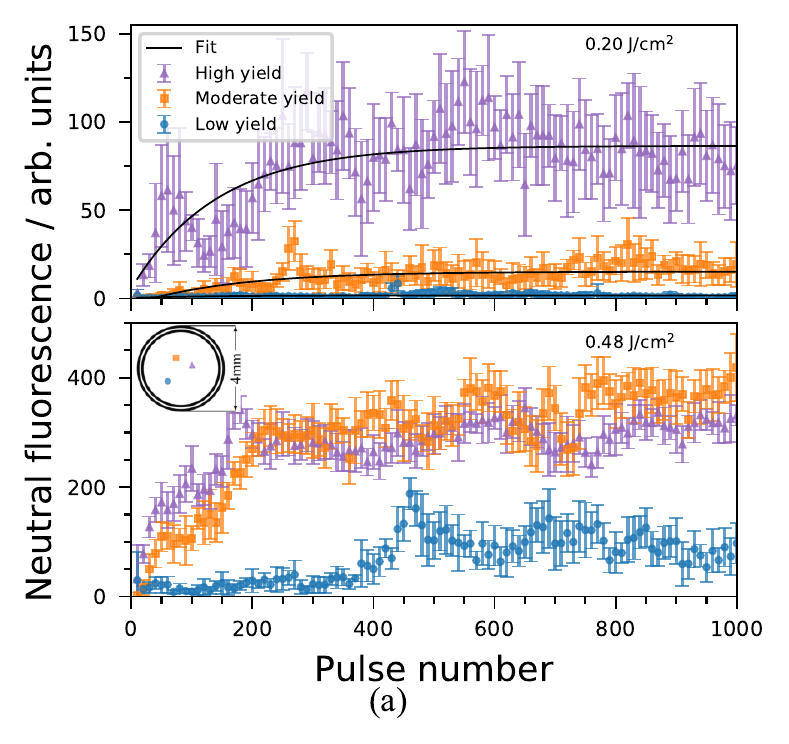}}
    	\vspace{-30px}
    	\hfill
    	\subfloat[labelformat=empty][\label{sfig:Spot-Lifetimes}]{\includegraphics[]{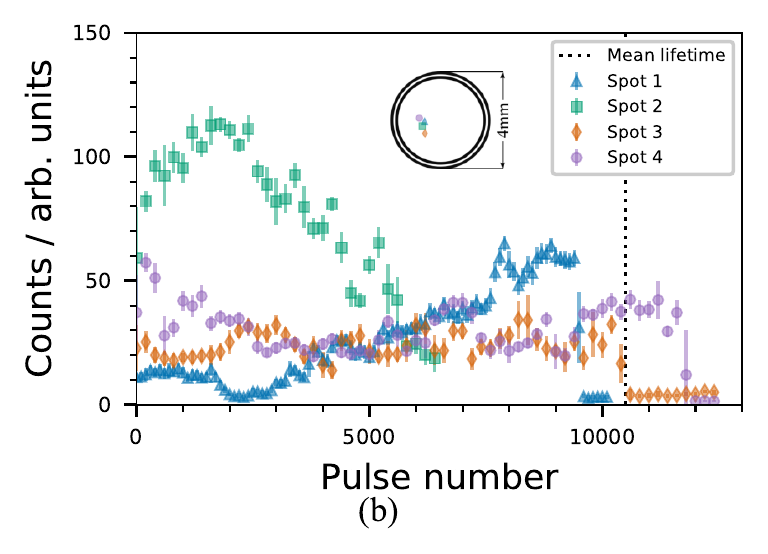}}
    	\caption{Ablation spot lifetimes and conditioning. First-step photoionization laser power is $\sim 150\times P_{sat}$. 
    		(a) Low pulse energy at $\SI{0.20}{\joule\per\centi\meter^2}$ (top), conditioning at $\SI{0.48}{\joule\per\centi\meter^2}$. 
    		At low pulse energy, some spots were high-yield, giving over 50 counts. 
    		Most spots were moderate-yield, giving over 10 counts
    		The remaining spots were low-yield, giving counts very near background.
    		(b) Neutral fluorescence spot lifetime for several spots.}
    	\label{fig:Lifetime-Conditioning}
    \end{figure}
    The conditioning process has highly variable outcomes (see Supplementary Materials), but three common behaviors emerged, illustrated in Figure \ref{fig:Lifetime-Conditioning}(a). 
    In these experiments, each fresh spot was exposed alternately to 10 conditioning pulses ($\SI{0.48}{\joule\per\centi\meter^2}$) and 10 low-fluence pulses ($\SI{0.20}{\joule\per\centi\meter^2}$).
    The neutral fluorescence at low and conditioning fluences is plotted against the number of conditioning pulses for three spots exhibiting the most common behaviors. 
    Out of seventeen fresh spots near the target center(see Supplementary Materials for the remaining data sets), five ($29\%$) were high-yield, with over 50 counts, four ($24\%$) were moderate-yield, with 10-50 counts, and eight ($47\%$) were low-yield, never exceeding 10 counts. 
    Typically, neutral-atom flux at low fluence plateaus after approximately 200 conditioning pulses.
    In a separate set of experiments, we observe that fluences below the threshold for ion production will not condition a fresh spot to produce neutral atoms even after several thousand pulses. 
    We also observe that spots towards the edge of the target are much less likely to yield significant neutral-atom flux (see Supplementary Materials).
    
    Even for high-yield spots, there is significant temporal variation in the ablated atomic flux, with the spot eventually giving no observable neutral-fluorescence signal.
    Therefore, an important metric is the lifetime of a single spot, i.e., the amount of pulses before the neutral-fluorescence signal reduces to near background. 
    As shown in Figure \ref{fig:Lifetime-Conditioning}(b), using a fluence of $\SI{0.15}{\joule\per\centi\meter^2}$ and a pulse repetition rate of $\SI{2}{\hertz}$, the average spot lifetime is $\sim 10,000$ pulses.
    Often, reconditioning a used-up spot revives it, but the lifetime after reconditioning is typically under $2,000$ pulses.
    
    On the low-volume target, a much higher fluence of $\SI{0.6}{\joule\per\centi\meter^2}$ is needed before neutral fluorescence is observable. 
    This signal diminishes to background levels after just tens of pulses (see Supplementary Material for details).
    Contrary to the natural-abundance barium target, reconditioning an area where the atomic flux is reducing never restores it, but instead speeds up the reduction.
    We interpret these observations, made for several large sweeps, as indications that the ablation completely depletes the source of neutral barium atoms. 
    This is supported by observations with similarly-prepared targets that laser ablation causes thin layers of $\mathrm{BaCl}_2$ to fragment and flake off their substrate material.
\section{Plume characterization}\label{sec:Ablation_Plume-Char}
    \begin{figure*}
    	\centering
    	\includegraphics[width=0.99\linewidth]{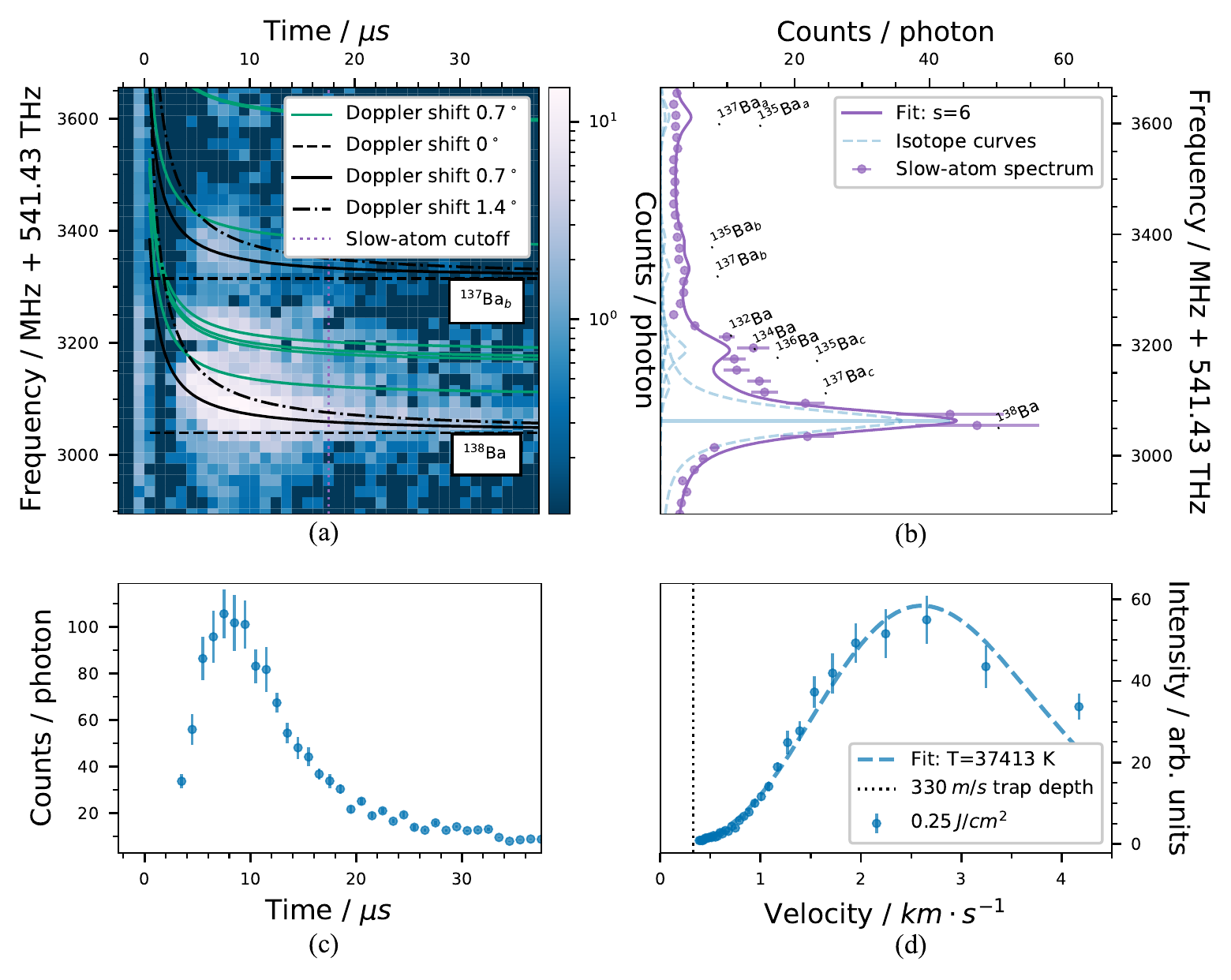}
    	\vspace{-15px}
    	\caption{Time-resolved spectroscopy. First-step photoionization laser power $12\times P_{sat}$, ablation fluence $\SI{0.25}{\joule\per\centi\meter^2}$. (a) Neutral fluorescence collected from our natural abundance target for different first-step photoionization frequencies and time-window starts. 
    		Two spots on the target were used for the data sets, with the second spot used for $f\geq\SI{3315}{MHz}$. 
    		Doppler shifts for a range of angles \SIrange[range-units = single]{0}{1.7}{\degree} are shown (relevant isotopes are black and others are green), with the fit curves as solid lines. 
    		The dotted-purple line denotes the boundary for slow atoms used to generate (b). 
    		(b) First-step photoionization spectrum from integrating over the times for slow atoms in (a) \cite{Baird1979, Bekk1979, Yamada1988}. 
    		(c) Overall time-of-flight distribution from integrating over all frequencies in (a). 
    		(d) Velocity distribution generated by scaling the time-of-flight data in (c) appropriately and converting to velocity (see text). 
    		The dotted line indicates the highest velocity an atom can be trapped with, based on our estimation of trap depth (see Supplementary Materials).}
    	\label{fig:Ablation-TOF-Spec}
    \end{figure*}
    Several characteristics of the ablated plume of neutral-barium atoms are illuminated by performing a time-resolved spectroscopy experiment \cite{Guggemos2015}, in which neutral-atom fluorescence is collected in $\SI{1}{\micro\second}$ time bins for different first-step photoionization laser frequencies. 
    From this data, we observe the Doppler shift of ablated atoms, generate a first-step photoionization spectrum showing resolved isotope peaks, a generate a velocity distribution.
    
    The time-resolved spectrum is shown in Figure \ref{fig:Ablation-TOF-Spec}(a).
    In this experiment, each point is the average of neutral fluorescence from five pulses, and each measurement is calibrated by a reference neutral-fluorescence measurement before and after it to control for the temporally varying signal. 
    For the calibration steps, the laser frequency is set to the $^{138}\mathrm{Ba}$ peak, and a $\SI{55}{\micro\second}$ time window is used, starting at $\SI{2.5}{\micro\second}$ after the ablation pulse.
    Photons collected in time bins closer to the moment of the laser pulse come from faster atoms in the ablation plume. 
    
    Only slow atoms with a lower kinetic energy than the trap depth can be confined in the Paul trap. 
    From the full time-resolved spectrum, the first-step photoionization spectrum for slow atoms is generated as shown in Figure \ref{fig:Ablation-TOF-Spec}(b). 
    Most isotopes' peaks are distinguishable, including the well-isolated $^{137}\mathrm{Ba}\:(j'=3/2)$ peak we use for trapping this isotope.
    A fit is done using the known theoretical-transition frequencies and transition strengths of different barium isotopes \cite{Bekk1979, Baird1979, Wijngaarden1995} and the isotopic abundances. 
    
    Integrating the time-resolved spectrum over frequency, results in the time-resolved distribution in Figure \ref{fig:Ablation-TOF-Spec}(c). 
    Using the fixed distance from the ablation target to the trap center, $d=\SI{14.6}{\milli\meter}$, and scaling by $1/\tau$, where $\tau$ is the time after the ablation pulse (to account for the transit time through the imaging-system field-of-view), results in the velocity distribution in Figure \ref{fig:Ablation-TOF-Spec}(d). 
    A Maxwell-Boltzmann distribution,
    \begin{equation}
    	\label{eq:TOF_MB}
    	f_v(\tau) = \frac{A}{\tau}e^{-\frac{md^2}{2\tau^2k_bT}},
    \end{equation}
    is fit to extract a plume temperature of $37,000~\SI{}{\kelvin}$. 
    Here, $A$ is an amplitude-scaling parameter, $v$ is atom velocity, $m$ is the average mass of natural-abundance barium, and $k_B$ is the Boltzmann constant. 
    The peak velocity of the distribution is $2,600~\SI{}{\meter\per\second}$. 
    
    If the first-step photoionization laser were perfectly perpendicular to the ablation plume, one would expect the laser frequency to have a global influence on the brightness in the time-resolved spectroscopy data. 
    In the case where the two are an angle $\theta$ off from perpendicular, high-velocity atoms see the laser frequency Doppler shifted. 
    Therefore, the peak frequency for different speeds of ions are Doppler shifted as follows:
    \begin{equation}
    	\label{eq:TOF_Doppler-Shift}
    	f_{ob} = f_s\left(1 + \frac{d\sin\theta}{c\tau}\right),
    \end{equation}
    where $f_s$ is the laser frequency and $c$ is the speed of light. A two-dimensional fit to the time-resolved spectrum is done using this model in conjunction with Equation \ref{eq:TOF_MB} and the model used to describe the spectrum in Figure \ref{fig:Ablation-TOF-Spec}(b) (see Supplementary Material). 
    The Doppler shifts resulting from this fit are shown as the black and green curves in Figure \ref{fig:Ablation-TOF-Spec}(a), with a fit angle of $\theta = \SI{0.7}{\degree}$.
    For this data set, the specific spots on the ablation target were $\sim\SI{200}{\mu m}$ from the target center. 
    Assuming the ablation target and first-step photoionization laser are well aligned to the viewport axes, the angle between the atomic beam and the laser is $\sim \SI{1}{\degree}$, very near the fit result of $\SI{0.7}{\degree}$. 
\section{Ion loading and selectivity}\label{sec:Iso-Sel}
    With a reliable source of neutral barium, it is straightforward to photoionize and confine ions within the trap.
    To pick a specific isotope of barium, the $\SI{554}{\nano\meter}$ laser frequency is parked on the desired peak in figure \ref{fig:Ablation-TOF-Spec}(b). 
    The $\SI{405}{\nano\meter}$ laser then fully ejects a valence electron of the selected isotope, resulting in an ion confined within the pseudo-potential well of our trap.
    If the kinetic energy of the ionized barium is below the trap depth potential, it will remain within the well.
    Cooling laser beams at $\SI{493}{\nano\meter}$ and $\SI{650}{\nano\meter}$, with frequency configuration depending on the chosen isotope, cool newly-confined ions until they become crystallized within a small area at the center of the trap.
    In this section, we present results for the loading rates and selectivities for both $^{138}\mathrm{Ba}^+$, and the less-abundant isotope, $^{137}\mathrm{Ba}^+$. 
    Finally, we discuss results from our low-volume barium-salt target.
    
    We define the loading rate as the average number of ions trapped for each ablation pulse.
    Using a first-step photoionization-laser saturation of $s=150$, the loading rate for $^{138}\mathrm{Ba}^+$ is measured to be very close to $\SI{1}{ions\per pulse}$ (20 pulses).
    Using the same parameters, the $^{137}\mathrm{Ba}^+$ loading rate is measured to be $\SI{0.1}{ions\per pulse}$ (53 pulses), very close to the expected loading rate based on comparing isotopic abundances and transition strengths between the two isotopes. 
    
    A dual-band-pass filter is used in our imaging system, allowing for measurement of both neutral-atom fluorescence during ablation, and ion fluorescence of trapped ions.
    We observe that the neutral-fluorescence signal of slow atoms, measured after the slow-atom cutoff in figure \ref{fig:Ablation-TOF-Spec}(a), is well correlated with the loading rate (see Supplementary Material). 
    Using this correlation, the expected loading rate can be actively monitored without needing to perform a statistical trapping experiment.
    
    \begin{figure}
    	\centering
    	\includegraphics[width=\linewidth]{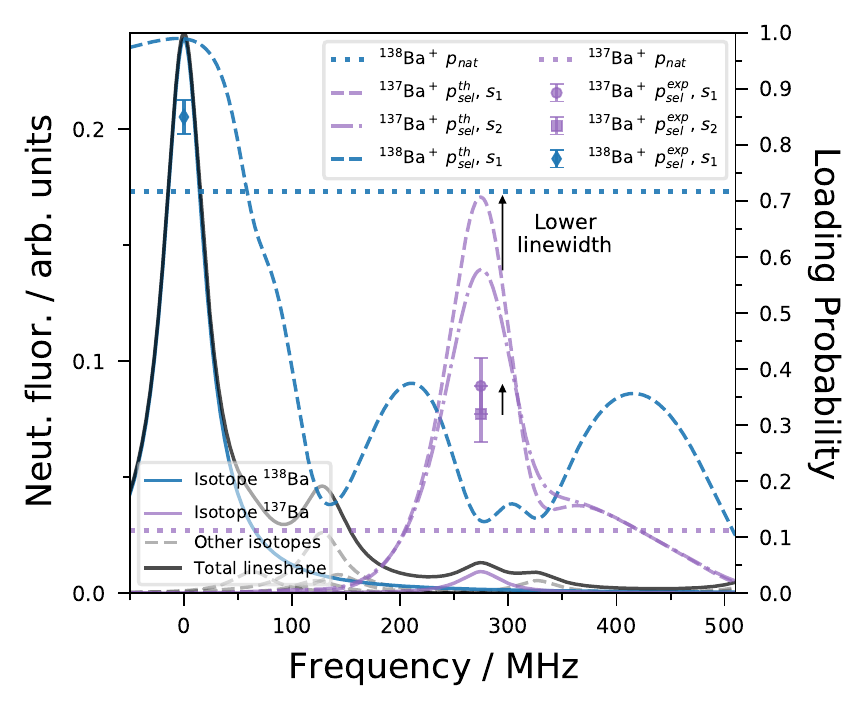}
    	\caption{
    		Selectivity of $^{137}\mathrm{Ba}^+$. 
    		Theoretical loading probabilities for isotopes of barium using two-step photoionization, $p_{sel}^{th}$, are calculated by comparing their Gaussian curves to the total lineshape.
    		Experimentally measured $p_{sel}^{exp}$ are shown as points.}
    	\label{fig:Selectivity}
    \end{figure}
    To quantify the ability to choose isotopes using the two-step photoionization method, we define the selectivity, $p_{sel}$, as the probability of loading an isotope. 
    We also define $p_{nat}$ as the natural selectivity, given by the isotope's natural abundance.
    We further define the selectivity enhancement as
    \begin{equation}
    	\varepsilon_{sel} = \ln{\left(\frac{p_{sel}}{1-p_{sel}}\right)} - \ln{\left(\frac{p_{nat}}{1-p_{nat}}\right)}.
    \end{equation}
    Here, each natural log is the logit function.
    This measure compares an achieved selectivity to the selectivity for a unbiased loading method, which traps according to an isotopes' $p_{nat}$.
    In our setup, we have the ability to discriminate between all naturally-abundant isotopes of barium except for $^{135}\mathrm{Ba}^+$.
    In order to measure the percent of desired ions loaded, considering all trapped atomic or molecular ions, chains of  $^{137}\mathrm{Ba}^+$ are trapped and the number of bright $^{137}\mathrm{Ba}^+$ ions is compared to the number of dark ions. 
    In Figure \ref{fig:Selectivity}, the expected theoretical selectivities for $^{138}\mathrm{Ba}^+$ and $^{137}\mathrm{Ba}^+$ are shown as a function of the first-step photoionization laser frequency, and for different saturations of the transition (i.e. different transition linewidths).
    The experimentally-measured selectivies are also shown in this Figure, with the $^{137}\mathrm{Ba}^+$ selectivity enhancement increasing from $\varepsilon=1.3$ to $\varepsilon=1.5$ by lowering the transition linewidth, and $^{138}\mathrm{Ba}^+$ giving a selectivity enhancement of $\varepsilon=0.8$. 
    In all cases, the achieved selectivity is lower than the theoretical. However for $^{137}\mathrm{Ba}^+$, the selectivity does improve with a lower transition linewidth as expected.
    
    No isotope-selective loading experiments were performed with the low-volume ablation target, owing to the inability to reliably produce neutral atoms.
\section{Conclusions}\label{sec:Conclusions}
    Many of the observations about the varying neutral-fluorescence signal can be explained by the result that the majority of atomic flux is directed perpendicular to the surface of the ablation target \cite{Laska2002, Thestrup2002, Laska2008}.
    Due to this effect, if the surface normal vector of a spot on the target is unsuitably shaped, atomic-flux density directed through the trap center is expected to be lower.
    We hypothesize that conditioning a spot or an area changes the surface profile, transforming the shape to be more suitably directed towards the trap center; that in addition to the common belief that conditioning serves to purge away a surface layer of contaminants \cite{Hendricks2007, Olmschenk2017}. 
    
    Others groups have opined that temporal reduction of neutral-fluorescence signal is the result of an extreme-pitting process, where the severely-angled ablation laser has dug a hole into the target such that the ablated spot no longer has line-of-sight to the center of the trap \cite{Knight1981, Chrisey1994, Leibrandt2007}.
    However, reference \cite{Hendricks2007} showed that for calcium-metal targets, the depth of pitting from ablation using low-pulse fluence is negligible. 
    Since we see no visible difference in a used spot on our target, we instead suspect the temporal reduction is due to a less severe surface-profile change.
    We see further evidence towards this explanation (see Supplementary Materials), but whether these effects are from a changing surface geometry is uncertain at this point. 
    
    The time-resolved spectroscopy results indicate that plume temperatures from ablating the barium-salt targets are high, at around $37,000~\SI{}{\kelvin}$. 
    Most other groups have measured a neutral atom ablation plume temperature of $1,000\mathrm{~to~}10,000~\SI{}{\kelvin}$ \cite{Sheridan2011, Guggemos2015, Vrijsen2019}, but some have measured similar temperatures as us \cite{Matos2014, Guggemos2015}.
    Our barium-salt target could be exhibiting different plume dynamics, or the contrast may partly be explained by a difference in data analysis.
    In order to take into account Doppler shift, a time-resolved distribution should be collected over a spectrum, as done in this paper and Reference \cite{Guggemos2015}.
    If the distribution is only collected for one frequency, and Doppler shift is present, the velocity distribution will appear to give a lower temperature.
    
    As shown in Figure \ref{fig:Selectivity}, we were unable to achieve the expected selectivity for $^{137}\mathrm{Ba}^+$.
    One mechanism that may be affecting selectivity for this isotope is charge-exchange \cite{Mortensen2004, Lucas2004, Tanaka2005, Tanaka2007}. 
    When trying to trap more ions, an already selectively-trapped $^{137}\mathrm{Ba}^+$ ion can experience an electron-transfer collision
    \begin{equation}
    	^{137}\mathrm{Ba}^+ + ^{138}\mathrm{Ba} \rightarrow ^{137}\mathrm{Ba} + ^{138}\mathrm{Ba}^+,
    \end{equation}
    resulting in a trapped-$^{138}\mathrm{Ba}^+$ ion instead.
    Because of its relatively-large abundance, this charge-exchange is likely to result in more $^{138}\mathrm{Ba}^+$ trapped, reducing the selectivity of other isotopes.
    Mortensen et. al. \cite{Mortensen2004} demonstrated this effect strikingly, using it to swap out most of a large crystal of $^{44}\mathrm{Ca}^+$ ions with the much more abundant $^{40}\mathrm{Ca}^+$ isotope.
    
    During data collection, we calibrate the wavelength of the 553 nm laser to the first-step photoionization $^{138}\mathrm{Ba}^+$ peak before every experiment (including the selectivity experiment). This is done by performing a frequency scan, collecting slow-atom neutral fluorescence (with a collection window starting $\SI{17}{\micro\second}$ after the ablation pulse, with a $\SI{55}{\micro\second}$ width).
    The calibration helps alleviate slow drifts in the measured peak frequency, which we attribute to the wavelength meter used for locking the laser, but because of Doppler shift, the found peak may be up to $\SI{10}{\mega\hertz}$ from the actual peak.
    Another observation is that our selectivity enhancement for $^{138}\mathrm{Ba}^+$ was lower than for $^{137}\mathrm{Ba}^+$ (i.e. it was less selective).
    Since our Doppler cooling lasers cool all even isotopes while trapping $^{138}\mathrm{Ba}^+$, they may be easier to trap, reducing the selectivity enhancement compared to for $^{137}\mathrm{Ba}^+$. 
    
    A few different methods have been employed to improve selectivity further \cite{Lucas2004, Tanaka2007, Toyoda2001, Kitaoka2012, Borisyuk2017, Alheit1996, Hasegawa2000}. 
    These methods involve heating the unwanted isotope out of the trap using either the cooling lasers \cite{Lucas2004, Tanaka2007, Kitaoka2012, Toyoda2001, Borisyuk2017}, or by applying an additional rf to the trap, exciting the secular motion \cite{Alheit1996, Hasegawa2000}.
    Alternatively, a different photoionization path could be employed, with the first step of photionization exciting to the $6^3\mathrm{P}_1$ state using a $\SI{791}{\nano\meter}$ laser \cite{Graham2014, Steele2007, Wang2011}.
    This scheme has larger isotope shifts, and has already been used to selectively load $^{137}\mathrm{Ba}^+$ ions \cite{Wang2011}. 
    
    We demonstrated that the low-density radioactive target is unsuitable for producing neutral-barium atoms with laser ablation, owing to the short timescale on which spots are depleted. 
    Our observed lifetime for neutral-atom production is many orders of magnitude shorter than that measured in References \cite{Hucul2017,Christensen2020_thesis} for direct $\mathrm{Ba}^+$ ion production. 
    We further observe evidence that our lifetime for direct-ion production (described in the Supplementary Material) is significantly shorter than that seen in \cite{Hucul2017,Christensen2020_thesis}. 
    We observed in separate experiments that melting and recrystallizing the $\mathrm{BaCl}_2$ before ablation, which was done in \cite{Hucul2017,Christensen2020_thesis} but not in our work, improves the mechanical integrity of a $\mathrm{BaCl}_2$ layer; we hypothesize that a high-temperature treatment may extend the lifetime of direct-ion production from $\mathrm{BaCl}_2$, either by adhering the salt to its substrate more securely or by impregnating some of the barium atoms deeper into the substrate \cite{Christensen2020_thesis, Laughlin-2015}. 
    Future work is needed to develop a reliable source for neutral $^{133}\mathrm{Ba}$ atoms.
    \begin{acknowledgements}
    	This research was supported in part by the Natural Sciences and Engineering Research Council of Canada (NSERC) and the Canada First Research Excellence Fund (CFREF).
    \end{acknowledgements}
    
\appendix
\section{Sweeping and conditioning}
\begin{figure}
	\centering
	\subfloat[\label{sfig:Conditioning-Spots_Lifetimes}]{\includegraphics[width=0.9\linewidth]{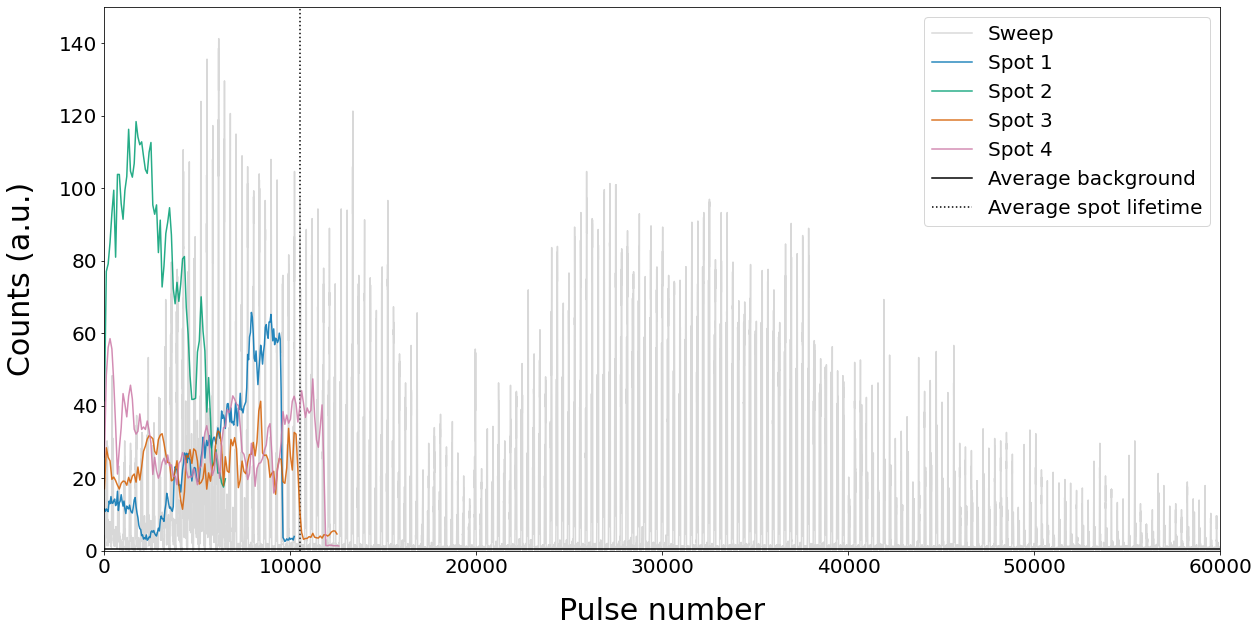}}
	\hfill
	\subfloat[\label{sfig:Conditioning-Sweep}]{\includegraphics[width=0.9\linewidth]{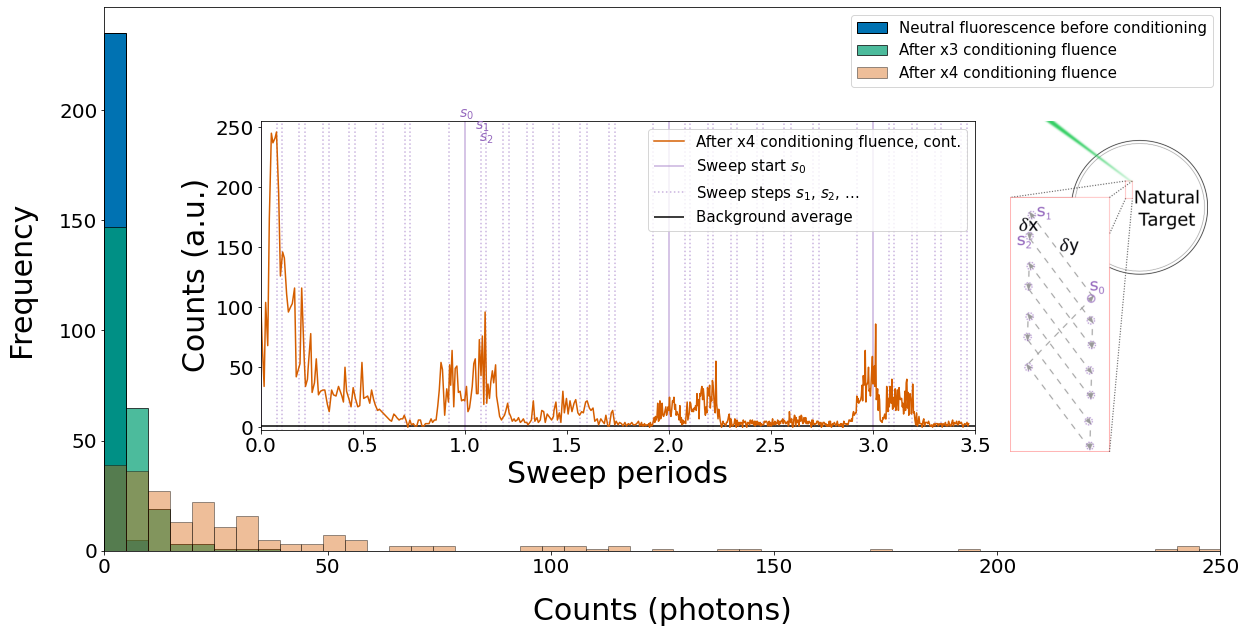}}
	\caption{
		Ablation lifetimes and conditioning. 
		The first-step photoionization laser power is set to $140\times P_{sat}$. 
		(a) Neutral fluorescence varies as a spot is ablated more, with an average overall lifetime of a spot being 10,500 pulses.
		Sweeping the laser across the target increases the amount of pulses before needing to move to $\sim 60,000$ pulses. 
		(b) Distribution of neutral fluorescence counts over a sweep area. As the area is conditioned, parts of the area generate more measured neutral fluorescence. 
		(inset) Raw sweep data after the area is conditioned. The signal varies significantly from sweep to sweep.}
	\label{fig:Conditioning}
\end{figure}

The ablation laser can be swept across the target over a conditioned patch using a servomotor-controlled mirror mount.
Other groups \cite{Hendricks2007, Sheridan2011} have done this to reduce the frequency of needing to find new spots, but we mostly use it to help characterize the spatial variation of neutral fluorescence signal across areas of the target.
We sweep the ablation laser as shown in the insets of figure \ref{fig:Conditioning}(b), with a fluence of $\SI{0.17}{\joule\per\centi\meter^2}$ and a repetition rate of $\SI{5}{\hertz}$.
The beam starts at position $s_0$, and sweeps over an area of $200\times 400\:\mu m$, returning to $s_0$ after $\sim\SI{60}{s}$.
The overall lifetime of such a sweep is $\sim 60,000$ pulses compared to $\sim 10,000$ for single spots, as shown in Fig. \ref{fig:Conditioning}(a). 
Variation within a single sweep period is apparent in the inset of Fig. \ref{fig:Conditioning}(b), and the explanation could be surface geometry - the surface profile could be directed away from the trap center in one part of the sweep, and well- directed in another. 
There's also considerable contrast sweep to sweep, a result which could be explained by ablation of slightly different spots each sweep. 

The results of preparing this area by conditioning are shown in Fig. \ref{fig:Conditioning}(b). 
After conditioning at various fluences above the base fluence, the neutral fluorescence signal increased to more than 50 times background in parts of the sweep.
Without the conditioning step, a fresh spot will not give appreciable neutral fluorescence signal at low fluence.
The ablation process is quite stochastic and multiple condition runs are frequently necessary.
Often, a spot never exhibits sufficient atomic flux to measure, possibly due to a poorly aligned surface profile.

\begin{figure*}
	\centering
	\includegraphics[width=\linewidth]{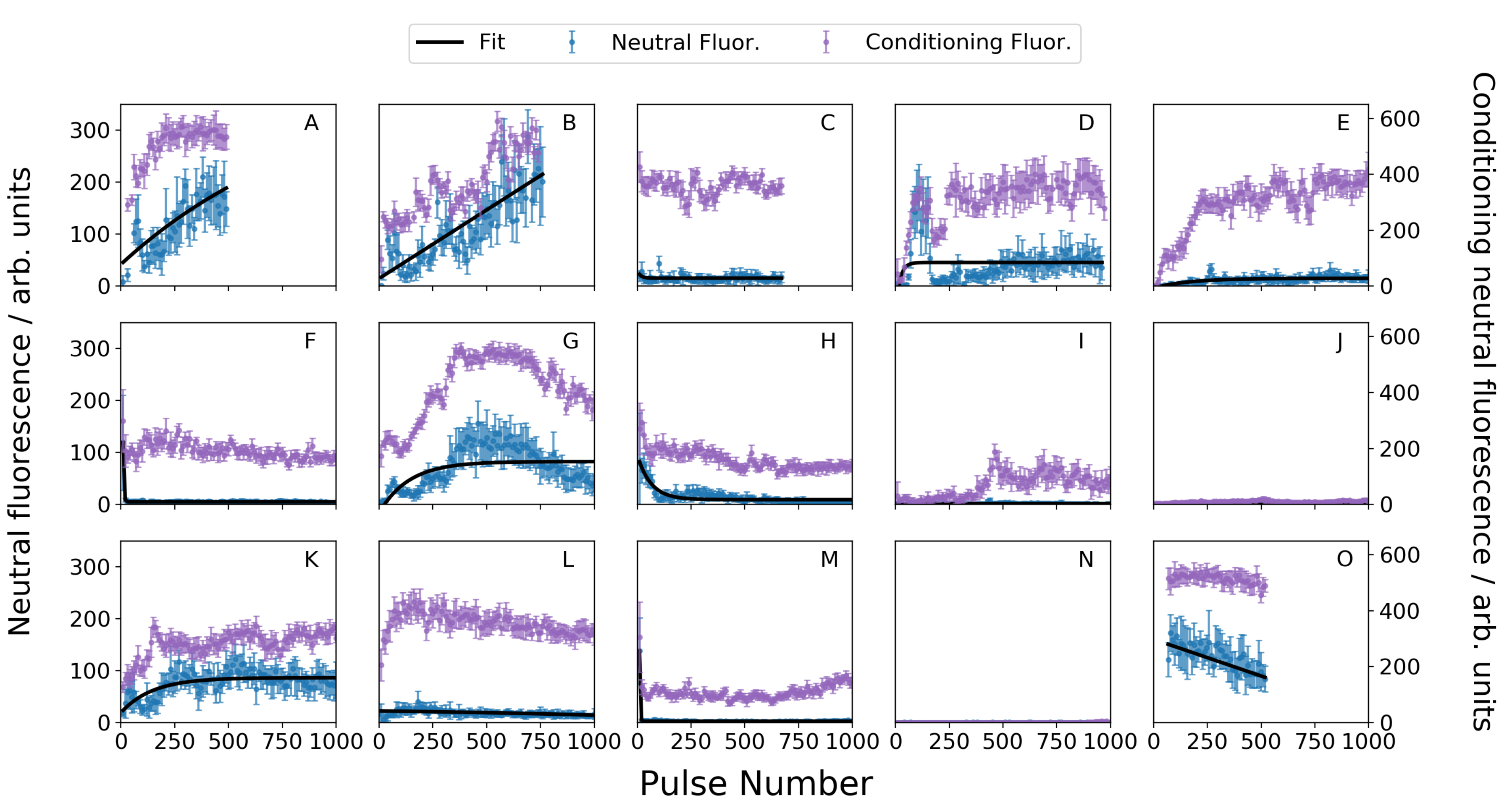}
	\caption{
		Conditioning spots near the center of the target. 
		Neutral fluorescence window start-time is set at $\SI{145}{\micro\second}$, and the integration width is set at $\SI{55}{\micro\second}$. 
		Neutral fluorescence varies from spot to spot, with all spots shown located in the central band of the ablation target (see target map in Figure 3).
		Conditioning at $\SI{0.48}{\joule\per\centi\meter^2}$ improved some spots' neutral fluorescence at low pulse energy ($\SI{0.2}{\joule\per\centi\meter^2}$) from background to 50 or more counts over several hundred pulses (spots A, B, G, K, O, described as high-yield in the main text). 
		For most spots, the neutral fluorescence counts are lower (spots C, D, E, L, moderate-yield), and for some, conditioning never results in sufficient atomic flux (spots F, H, I, J, M, N, low-yield). 
		Two additional low-yield spots were not shown here.}
	\label{fig:Conditioning-variability}
\end{figure*}

\begin{figure*}
	\centering
	\includegraphics[width=\linewidth]{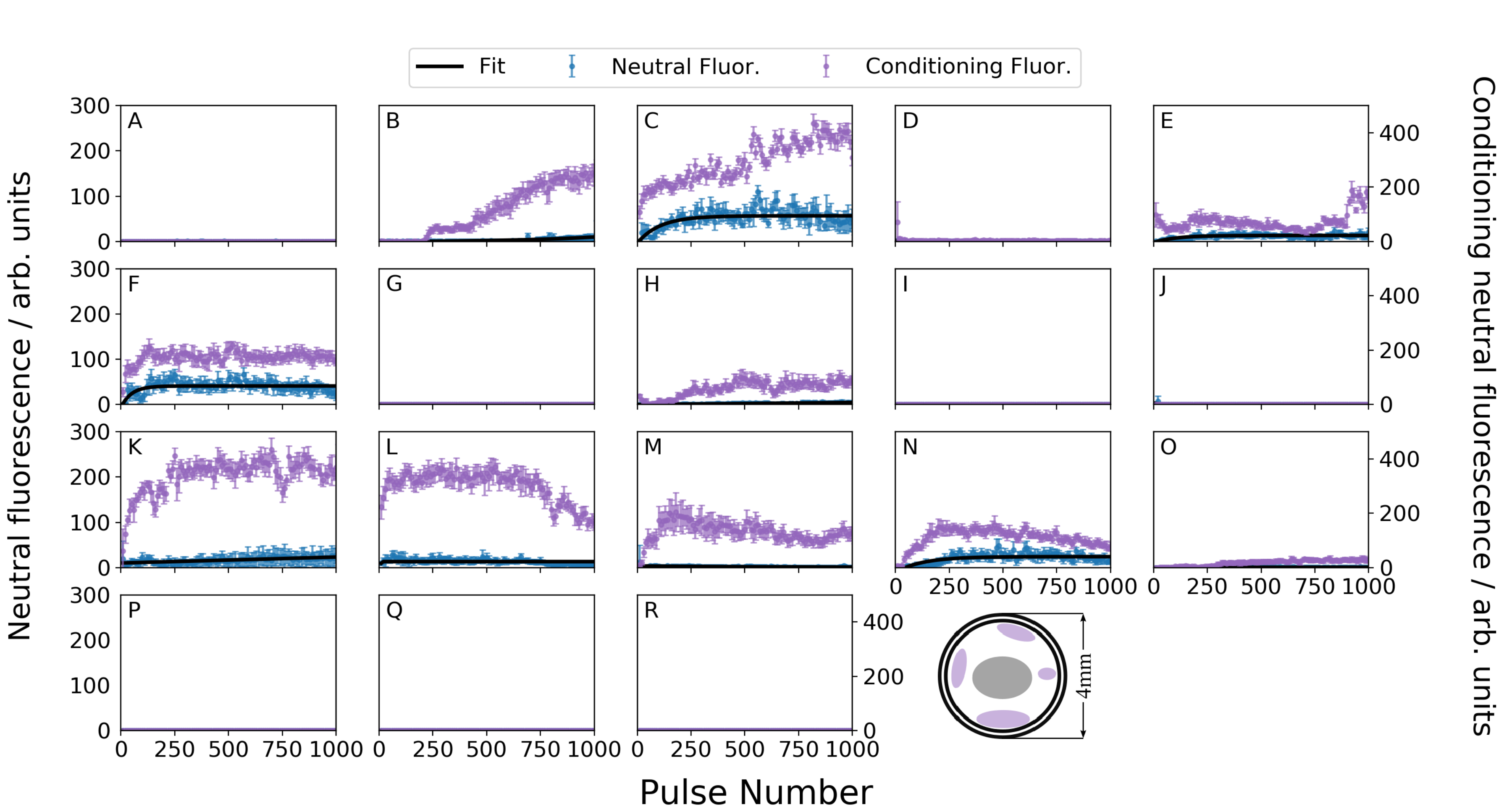}
	\caption{
		Conditioning spots near the edges of the target. 
		Neutral fluorescence varies from spot to spot, with all spots shown located around the perimeter or edges of the ablation target.
		Conditioning at $\SI{0.48}{\joule\per\centi\meter^2}$ improved some spots' neutral fluorescence at low pulse energy ($\SI{0.2}{\joule\per\centi\meter^2}$) from background to 10-50 counts (spots B, C, E, F, K, L, N, described as moderate-yield in the main text). 
		For most spots, conditioning never results in sufficient atomic flux (spots A, D, G, H, I, J, M, O, P, Q, R, described as low-yield). 
		The spots are distributed around the perimeter of the target as follows: spots A, D, E and F are located along the left perimeter, spots B, C, G, I, J are located along the bottom, spots H, K, L, M, P, R are located along the top perimeter and finally, spots R, O and Q are located on the right. A target map with these regions highlighted is included in Figure 3 for reference.}
	\label{fig:Conditioning-variability_edge}
\end{figure*}

There is a significant amount of variability amongst the outcomes of conditioning a single spot on the ablation target. In these experiments, each fresh spot was exposed alternately to 10 conditioning pulses ($\SI{0.48}{\joule\per\centi\meter^2}$) and 10 low-fluence pulses ($\SI{0.20}{\joule\per\centi\meter^2}$), and neutral-atom fluorescence was collected for each pulse. 	Out of seventeen fresh spots collected, fifteen of which are illustrated in Figure 2, five ($29\%$) were high-yield, with 50+ counts after several hundred conditioning pulses, four ($24\%$) were moderate-yield, with 10-50 counts, and eight ($47\%$) were low-yield, never exceeding 10 counts. These three common behaviours that emerged were illustrated in Figure 3 (a) of the main document. All seventeen spots were collected across the central region of the target.  

In Figure \ref{fig:Conditioning-variability}, the neutral fluorescence at low and conditioning fluences is plotted against the number of conditioning for each spot. Typically, the neutral-atom flux at low fluence stops increasing after approximately 200 conditioning pulses. Each data point was averaged across either of the 10 low-fluence or 10 conditioning pulses, respectively. Figure \ref{fig:Conditioning-variability_edge} shows the same experiment but for spots near the edges of the target. 
For these spots, finding a good spot was much harder than for those in the center area of the target.
Only 7 out of 18 spots gave even moderate-yield neutral fluorescence, and most spots didn't even give significant conditioning fluorescence.
For spots towards the left or right side of the target, this could be because of extra Doppler shift resulting from the ablation plume passing through the 553 beam at a non-perpendicular angle.
For spots towards the top or bottom, this reduction could be partly from obscuration by one of the trap rods.

\section{Finding a new spot}
The practice of actually conditioning a new spot so that we can use it for trapping is still somewhat indeterminate, owing to the many unknowns in the ablation process. 
A typical algorithm for preparing a new spot is:
\begin{enumerate}
	\item Pick a fresh spot on the target, unused for anything yet.
	\item Check for neutral fluorescence at low pulse fluence, fluence typically used to trap ~\SIrange[range-units = single]{0.15}{0.3}{\joule\per\centi\meter\squared}
	\item If signal is less than 50x background, increase to a conditioning pulse fluence of $\SI{0.6}{\joule\per\centi\meter\squared}$ and pulse 50-100 times
	\item Return to low pulse fluence and measure the neutral fluorescence
	\item If signal is higher than before, pulse several hundred times to see if it improves. Otherwise, move to a higher conditioning fluence and pulse 50-100 times
	\item Return to low pulse fluence and measure the neutral fluorescence
	\item Repeat this usually at least 2-3 times. If this spot doesn't exhibit significant neutral fluorescence, then move on to a new spot and repeat the process. Typically, a fresh spot should be at least a beam waist away from the previous spot
\end{enumerate}

While moving to a new spot more than a beam waist away from the previous spot often works, sometimes you must move by more 2 beam waists. 
As mentioned above, empirically, this should work around $30\%$ of the time. So you shouldn't have to try more than around 3 or 4 fresh spots before successfully preparing a spot for use.
\section{Ablation pulse timing}
\begin{figure}
	\centering
	\includegraphics[width=\linewidth]{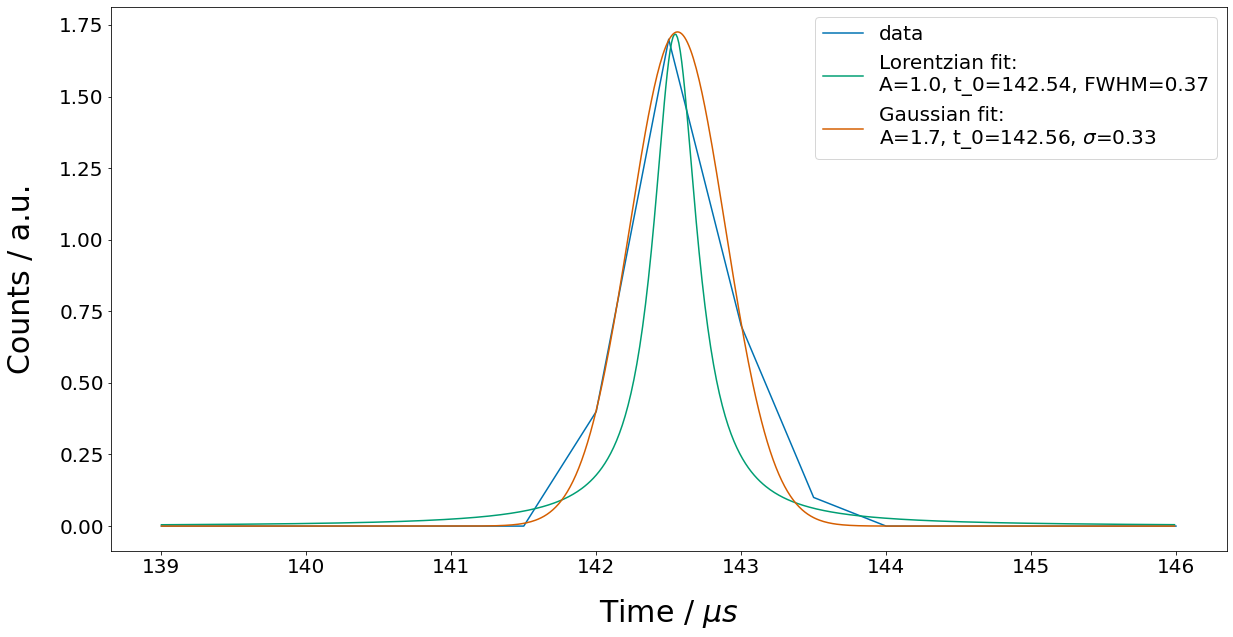}
	\caption{
		Ablation pulse timing.}
	\label{fig:Ablation-Timing}
\end{figure}
Upon triggering the ablation laser via a TTL signal, the flash lamp turns on. 
The q-switch trigger is then delayed by around $140 \mu s$ as the flash lamp ramps up. 
Within $0.1 \mu s$ after the q-switch, the $\sim 5 ns$ laser pulse is released. 
Ideally, the timing of this whole process would be consistent every time so that we accurately gather timing information about the ablation process.
However, in practice we see there is some amount of variation from shot to shot. 
Figure \ref{fig:Ablation-Timing} shows a time measurement of the 532 nm light after the flashlamp has been triggered.
Data is gathered at a resolution of 500 us, with the same integration time.
Using a Gaussian fit, we see that the ablation pulse happens at $142.5 \mu s$, with a standard deviation of $\sigma = 0.3 \mu s$.
We have tested our control system's TTL timing, and determined it to be precise to within nanoseconds.
Therefore, the inconsistency most likely lies within the laser, in the Q-switch delay timing. 
This timing variation manifests itself as a Doppler broadening in any TOF experiments.
\section{Time-resolved spectrum fitting}
\begin{figure*}
	\centering
	\subfloat[\label{sfig:TOF-Spect_2D_LowFluence}]{\includegraphics[height=0.4\linewidth, width=0.4\linewidth]{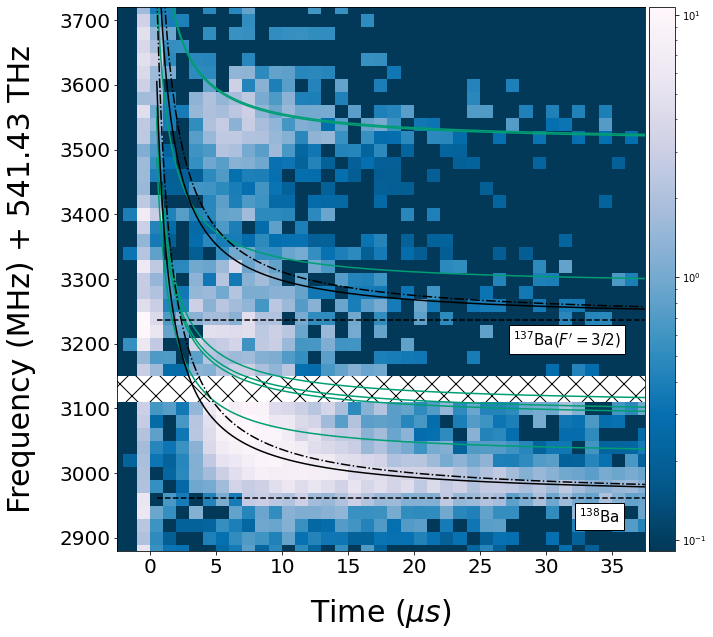}}
	\subfloat[\label{sfig:TOF-Spect_2D_Fit_LowFluence}]{\includegraphics[height=0.4\linewidth, width=0.4\linewidth]{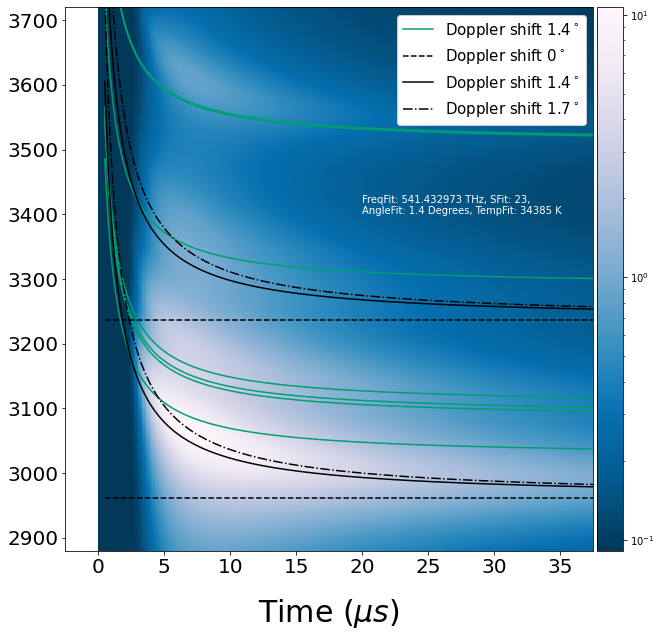}}
	\hfill
	\subfloat[\label{sfig:TOF-Spect_2D_HighFluence}]{\includegraphics[height=0.4\linewidth]{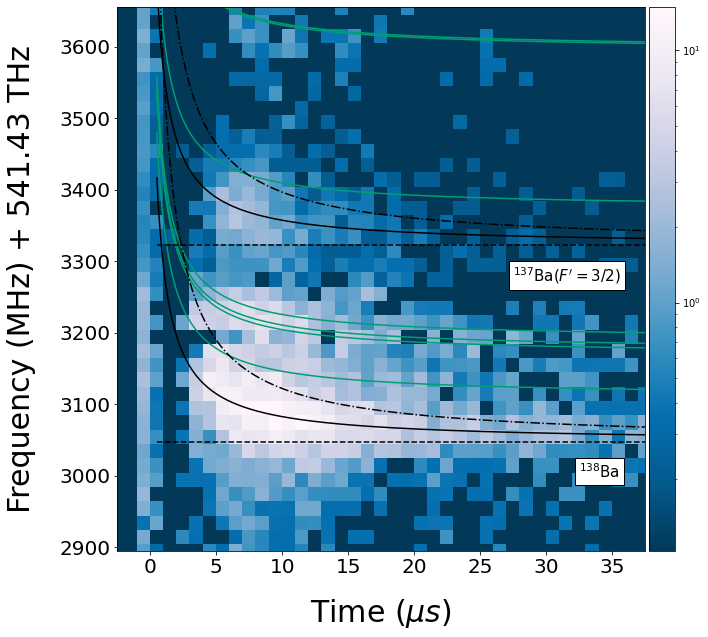}}
	\subfloat[\label{sfig:TOF-Spect_2D_Fit_HighFluence}]{\includegraphics[width=0.4\linewidth, width=0.4\linewidth]{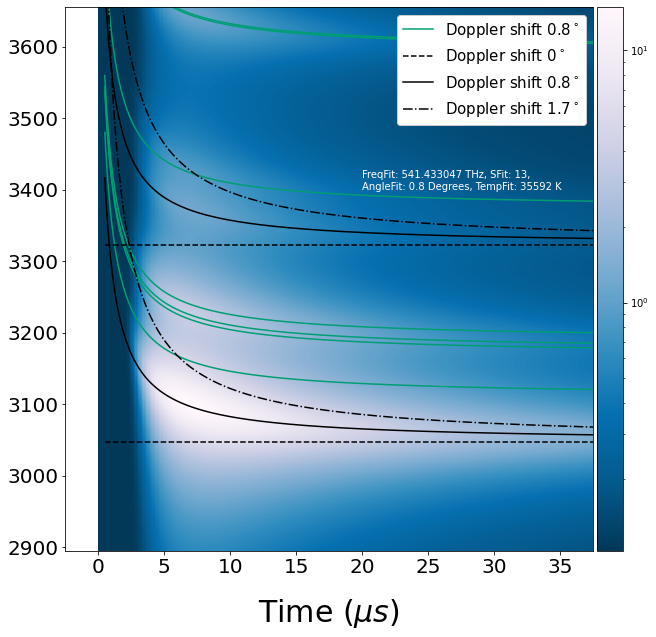}}
	\caption{
		TOF spectrum experiments. 
		(a, c) Raw 2D TOF spectrum data using $\SI{0.20}{\joule\per\centi\meter^2}$ and $\SI{0.25}{\joule\per\centi\meter^2}$ fluences respectively. 
		In (a), two different spots on the target were used for the data sets $f < 3120$ and $f \geq 3140$ respectively; wavelength calibration drifted by $\sim\SI{40}{MHz}$ in this time, leading to the gap in the data. 
		In (c), two different spots were also used for the data sets, with the second spot used for $f \geq 3315$. 
		(b, d) The corresponding 2D fits to the data.}
	\label{fig:Ablation-TOF-Spec_Fits}
\end{figure*}
Here, we present all of the time-resolved spectroscopy data, along with the 2D fits to the model presented in the text. Surprisingly, the lower fluence experiment resulted in a higher fitted temperature. This is likely due to the fit standard deviation being very high. 

\begin{figure}
	\centering
	\includegraphics[width=\linewidth]{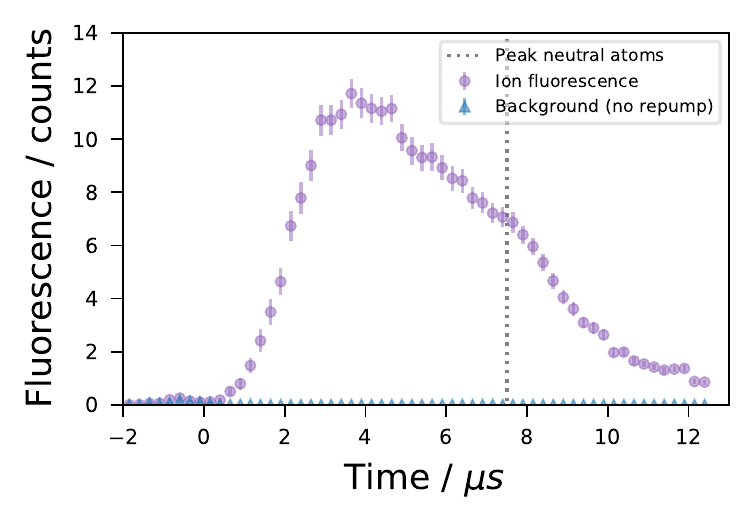}
	\caption{
		Ion fluorescence TOF. 
		The signal peaks at around $\SI{4}{\micro\second}$, compared to the neutral fluorescence signal which peaks at around $\SI{8}{\micro\second}$.}
	\label{fig:Ablation-TOF_Ion}
\end{figure}
We also present the result of a TOF experiment on the ion fluorescence in Figure \ref{fig:Ablation-TOF_Ion}. The speed of ions is nearly twice the speed of neutral atoms.
\section{Loading rate correlation with neutral fluorescence}
\begin{table*}[htb!]
	\centering
	\begin{tabular}{ccccccc}
		\toprule\toprule
		\textbf{Date} & \textbf{Fluence ($J/cm^2$)} & \textbf{Power ($\mu W$)} & \textbf{Time ($\mu s$)} & \textbf{Fluor.} & \textbf{Est. Fluor.} & \textbf{Load. Rate} \\\midrule
		11/26/2020 & 75 & 9 & 145 & 40 & 20 & 0.25\\
		& 75 & 90 & 145 & 40 & 40 & 0.31\\
		12/09/2020 & 49 & 80 & 145 & 100 & 50 & 0.09\\
		& 60 & 80 & 145 & 100 & 50 & 0.31\\
		& 60 & 80 & 145 & 500 & 25 & 0.07*\\
		12/14/2020 & 60 & 8 & 145 & 100 & 25 & 0.25\\
		02/17/2021 & 65 & 10 & 145 & 25 & 7.5 & 0.07$^\dagger$\\
		02/18/2021 & 65 & 10 & 145 & 25 & 7.5 & 0.05\\
		03/03/2021 & 140 & 80 & 160 & 100 & 100 & 1.0$^\circ$ \\
		03/04/2021 & 100 & 8 & 160 & 5-20 & 12 & 0.08\\
		& 140 & 80 & 160 & 120-240 & 170 & 1.0\\
		& 140 & 8 & 160 & 60-100 & 80 & 0.6\\\bottomrule
	\end{tabular}
	\label{tab:Neutral-Fluor-Correlation}
	\caption{
		Neutral fluorescence measurements and loading rates at different times and using different spots on the target. 
		The time column refers to the start of the fluorescence integration window. 
		Fluorescence is measured in counts per $\SI{55}{\micro\second}$ time window. 
		Loading rate is the number of ions trapped per pulse. Estimated fluorescence is an estimation of what each of the neutral fluorescence measurements would be if taken with $\sim\SI{10}{\micro\watt}$, and with a window time of $\SI{160}{\micro\second}$. 
		*This experiment was a sweep over an area, with counts of around 500 for only $10\%$ of the sweep. 
		$^\dagger$Newly calibrated wavemeter. 
		$^\circ$Dual pass filter added, giving us the ablility to measure neutral fluorescence while trapping.}
\end{table*}
We used neutral fluorescence as our measure to investigate the ablation process and the atomic plume.
This measure is further useful as an estimate of the loading rate.
Table \ref{tab:Neutral-Fluor-Correlation} shows the fluorescence counts and loading rate results from many different days and different spots on the target. 
After awhile, we decided to move the neutral fluorescence integration-window start time to $\SI{160}{\micro\second}$, since only the slower atoms are trappable, and to reduce the effect of Doppler shift. 
For all of this data, the total integration time is $\SI{55}{\mu\second}$.
To compare the different experiments, the neutral fluorescence measurements must be scaled based on how the parameters typically change them.
We found that reducing the ionization laser power from $\sim\SI{85}{\micro\watt}$ to $\sim\SI{8}{\micro\watt}$ decreased the neutral fluorescence signal to near half the original signal.
We found that moving the integration time start from $\SI{145}{\micro\second}$ to $\SI{160}{\micro\second}$ also reduced the signal by near half. 
Using these observations, we calculate the estimated neutral fluorescence if all experiments were using lower ionization power and a later integration time start.
\begin{figure}
	\centering
	\includegraphics[width=\linewidth]{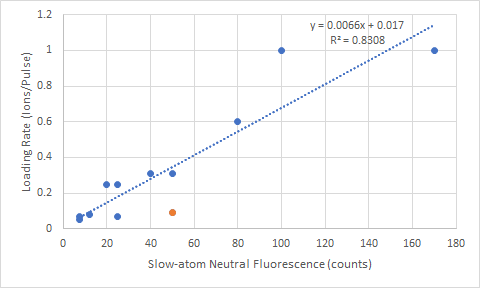}
	\caption{
		Loading rate vs neutral fluorescence for the experiments listed in Table \ref{tab:Neutral-Fluor-Correlation}. 
		The variables are very well correlated, with an $R^2 = 0.83$.
		The orange data point was excluded for the fit because it was done over a sweep in which spacial variation was high.}
	\label{fig:Fluorescence-Correlation}
\end{figure}

Figure \ref{fig:Fluorescence-Correlation} shows the loading rate vs estimated neutral fluorescence for all of the different experiments listed in Table \ref{tab:Neutral-Fluor-Correlation}.
With $R^2=0.83$, we see that the two variables are very well correlated.
We can therefore use neutral fluorescence as an estimate for loading rate without having to collect trapping statistics.
Initially, we were using two interchangeable single band-pass filters so that we could only measure either neutral fluorescence or ion fluorescence, not both at the same time.
Now, we use a dual-band pass filter which allows us to measure neutral fluorescence and then see the ion fluorescence after we've trapped an ion.
With this setup, it is possible to dynamically monitor the expected loading rate by simply measuring the neutral fluorescence every time the ablation laser is pulsed.

\section{Trap depth}
One can make a rough estimate for the trap depth of our four-rod Paul trap by using the results of direct-ion trapping, presented in section \ref{Direct-ion loading}.
From these experiments, we observed that ions begin to be trapped after the trap rf is switched on at $\sim 170 \mu s$, around $27 \mu s$ after the ablation pulse.
Next, an estimate trap turn-on time is needed. 
We have measured our trap's Q-factor to be $Q=350$, and we use an rf frequency of $\sim 20 MHz$ for trapping.
Therefore, the frequency width of the signal is $\sim 60 kHz$, giving an estimated turn-on time of $\sim 17 \mu s$.
So overall, after the laser is pulsed and ablation begins it takes $\sim 44 \mu s$ before atoms are trapped. 
Given the $14.6 mm$ distance to the trap center, the first atoms to be trapped in direct-ion loading have a velocity of $v < 330 m/s$.
That is, only atoms with a velocity slower than this are trappable. 
Given this estimate for the upper-bound of trappable barium atoms, we estimate our trap to have a depth of $0.08 eV$.
With this low trap depth, the estimated percentage of atomic flux that is trappable is under $1\%$.
\section{Loading rate vs cooling freq.}
\begin{figure}[H]
	\centering
	\includegraphics[width=\linewidth]{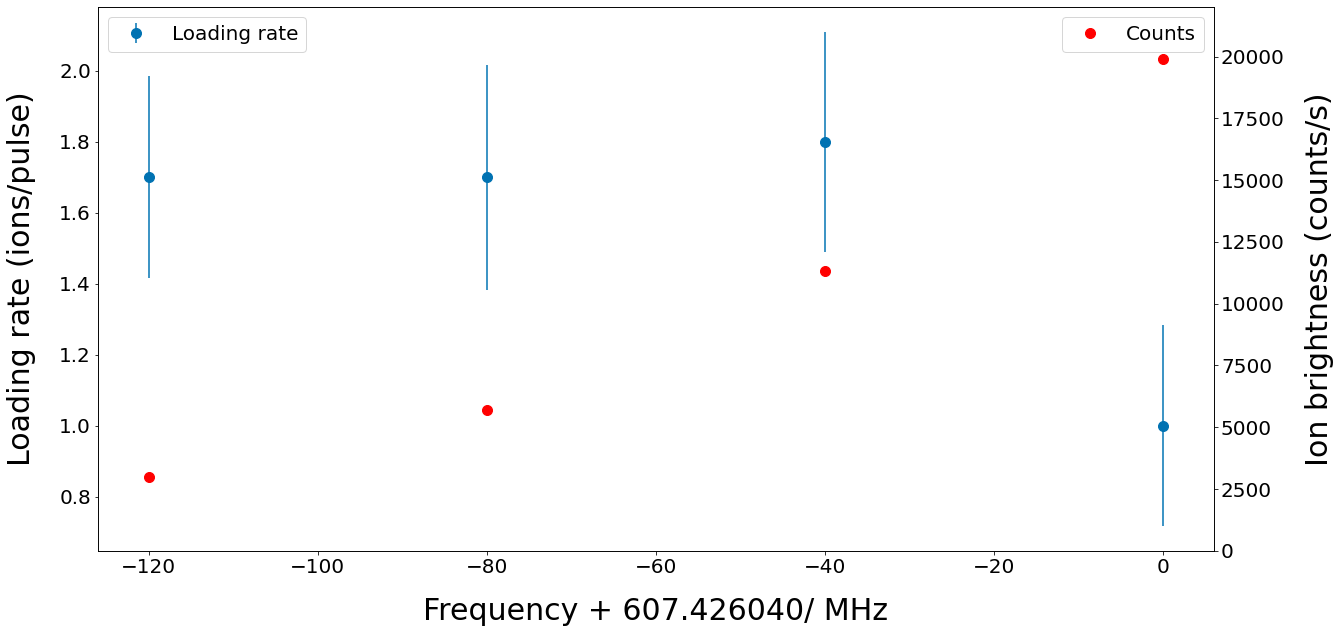}
	\caption{
		Loading rate for different cooling laser frequencies detuned from resonance. 
		Ion brightness for each frequency is also shown.}
	\label{fig:LoadingRate-vs-Cooling-Freq}
\end{figure}
As shown in Figure \ref{fig:LoadingRate-vs-Cooling-Freq}, loading ions is easier with a detuned cooling ($\SI{493}{\nano\meter}$) laser. 
However, with the laser on resonance, an already trapped ion fluoresces with near double counts than with $\SI{40}{\mega\hertz}$ detuning.
This indicates a better cooling rate for an already crystallized ion.
The advantage of detuning for loading rate can be seen when an uncrystallized or unconfined ion is considered: these hot ions experience a large Doppler shift, and will absorb red detuned light more than near-resonant light.
Therefore, using a detuning can confine and cool ions that otherwise would have escaped the trap.
In practice, during trapping we set our cooling laser $\SI{40}{\mega\hertz}$ red detuned from the optimal frequency.

\section{Selectivity vs pulse fluence}
\begin{figure}[H]
	\centering
	\includegraphics[width=\linewidth]{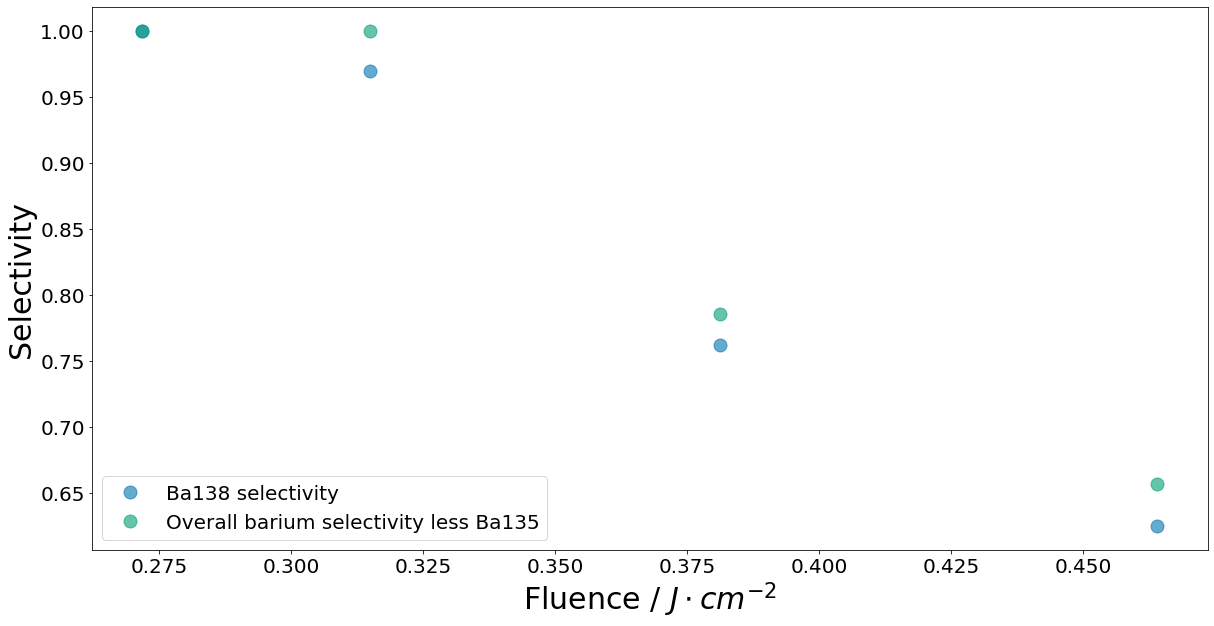}
	\caption{
		$^{138}\mathrm{Ba}^+$ and overall barium selectivity for different pulse fluences. 
		The setup is not able to discriminate $^{135}\mathrm{Ba}^+$.}
	\label{fig:Selectivity-vs-Fluence}
\end{figure}
An interesting observation is that the selectivity for $^{138}\mathrm{Ba}^+$ decreases dramatically with too high a pulse fluence.
As shown in Figure \ref{fig:Selectivity-vs-Fluence}, we observe that this reduction begins at very near the onset of direct-ion production.
In this experiment, we were able to discriminate between all barium isotopes but $^{135}\mathrm{Ba}^+$ (natural abundance: 6\%).
So in this Figure, we also present the likelihood that a trapped ion was a barium ion (of some other naturally occurring isotope besides $^{135}\mathrm{Ba}^+$).
As shown, most of the non $^{138}\mathrm{Ba}^+$ ions trapped were not even other barium isotopes.
Given the correlation of this relationship with the ion production, we hypothesize that these dark ions are not barium at all.
We believe these dark ions are some other atom or molecule being generated by the ablation.

\section{Direct-ion loading}
As a comparison, we also employed a direct ion-loading method similar to the one described by Hucul et. al. \cite{Hucul2017}, who use an ablation target similar to our low-volume target. 
Direct-ion loading is performed by trapping barium ions that are generated directly from the laser ablation process. First, the RF voltage to the ion trap is kept off to remove the potential barrier. Then, an ablation laser pulse of sufficient fluence to produce barium ion directly ($\sim\SI{0.5}{\joule\per\centi\meter^2}$) is sent to a barium chloride target. The trap RF voltage is then turned on after an appropriate time delay to capture the ions with kinetic energy below the potential barrier. Lasers of wavelengths $\SI{493}{\nano\meter}$ and $\SI{650}{\nano\meter}$ are then used to laser cool the trapped ion(s).

To determine the appropriate time delay for the RF voltage turn-on time, the loading efficiency is investigated for different turn-on time delays for the trap RF voltage. The experiment is performed on the high-volume target as it has a more consistent loading rate.
\begin{figure}[H]
	\centering
	\includegraphics[width=\linewidth]{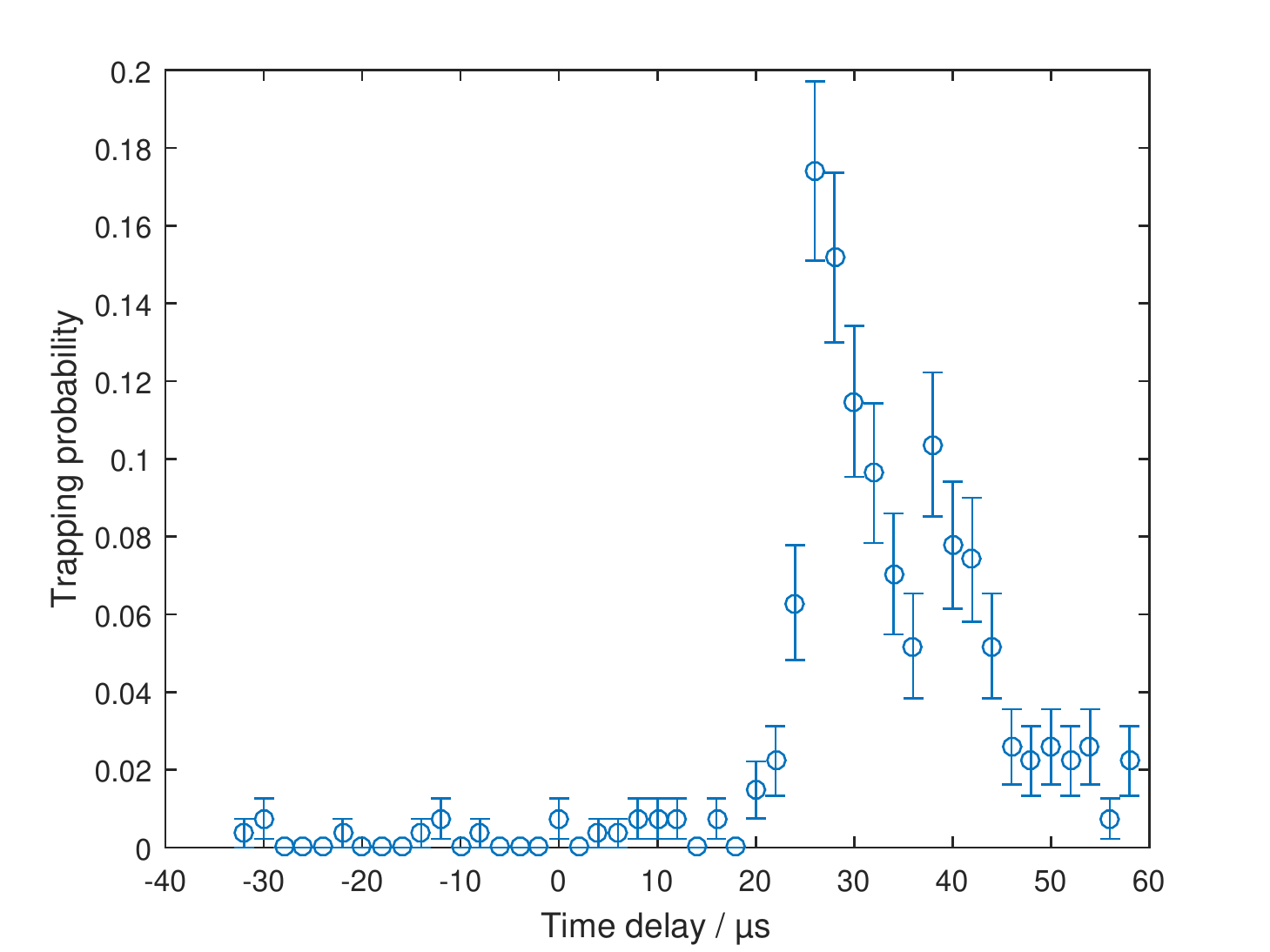}
	\caption{
		Trapping probability of $^{138}\mathrm{Ba}^+$ from the high-volume target using the direct-ion loading method. 
		The horizontal axis denotes the time delay of the trap RF voltage turn-on time from the moment the ablation laser pulse reaches the target. 
		The sample size for each data point is 270. 
		The ablation fluence used is $\SI{0.32}{\joule\per\centi\meter\squared}$.}
	\label{fig:Direct-ion_trapping_prob}
\end{figure}

From Figure \ref{fig:Direct-ion_trapping_prob}, the direct-ion trapping probability is highest when the trap RF voltage is turned on $\SI{26}{\micro\second}$ after the ablation laser pulse reaches the target.

We also investigated the loading efficiency of the direct-ion loading method compared to the two-step photoionization method. A pulse fluence of $\SI{0.32}{\joule\per\centi\meter\squared}$ is used to ablate on a single spot on the high-volume target. The $\SI{554}{\nano\meter}$ laser frequency is set to the fluorescence peak of neutral $^{138}\mathrm{Ba}$. The $\SI{493}{\nano\meter}$ and $\SI{650}{\nano\meter}$ are set to cool singly charged $^{138}\mathrm{Ba}^+$ ions. First, 300 loading attempts are performed using the two-step photoionization method. Then, another 300 loading attempts are performed using the direct-ion loading method. The trap RF voltage is turned on $\SI{27}{\micro\second}$ after the ablation laser hits the target. Lastly, another 300 loading attempts are performed without the two-step photonioization lasers nor timing the trap turn on time. This last set of loading attempts serve as a control experiment. We make the assumption that the flux density produced by ablation is consistent throughout these experiments in order to make direct comparisons of the loading efficiencies of the different loading methods. The loading efficiencies are found to be $15.7 \%$, $39.0 \%$ and $1.7 \%$ respectively for the two-step photoionization method, the direct-ion loading method and the control experiment respectively.

\section{Low-volume target}
We attempted to trap barium ions using the two-step photoionization method at various fluences from \SIrange[range-units = single]{0.6}{3.8}{\joule\per\centi\meter^2}.
For a fresh ablation spot, we trap $^{138}\mathrm{Ba}^+$ (which has an abundance of 53.3\% on the low-volume target) less than 10 times before being unable to trap, with a loading rate of $\sim \SI{0.05}{ions\per pulse}$ (much lower than from our high-volume target).

To investigate the reliability of an ablation spot on this target, we checked for neutral barium fluorescence rate as we ablate the target. 
The ablation laser is swept along a line on the low-volume target, ablating along that line. 
For each ablation pulse, the neutral barium fluorescence from the $\SI{554}{\nano\meter}$ is collected on a PMT with an integration time of $\SI{100}{\milli\second}$. 
The pulse fluence used is $\SI{0.59}{\joule\per\centi\meter\squared}$.

\begin{figure*}
	\centering
	\subfloat[\label{sfig:RadNeutralFluorescence_125uJ_beforehighpower}]{\includegraphics[
		width=0.4\linewidth]{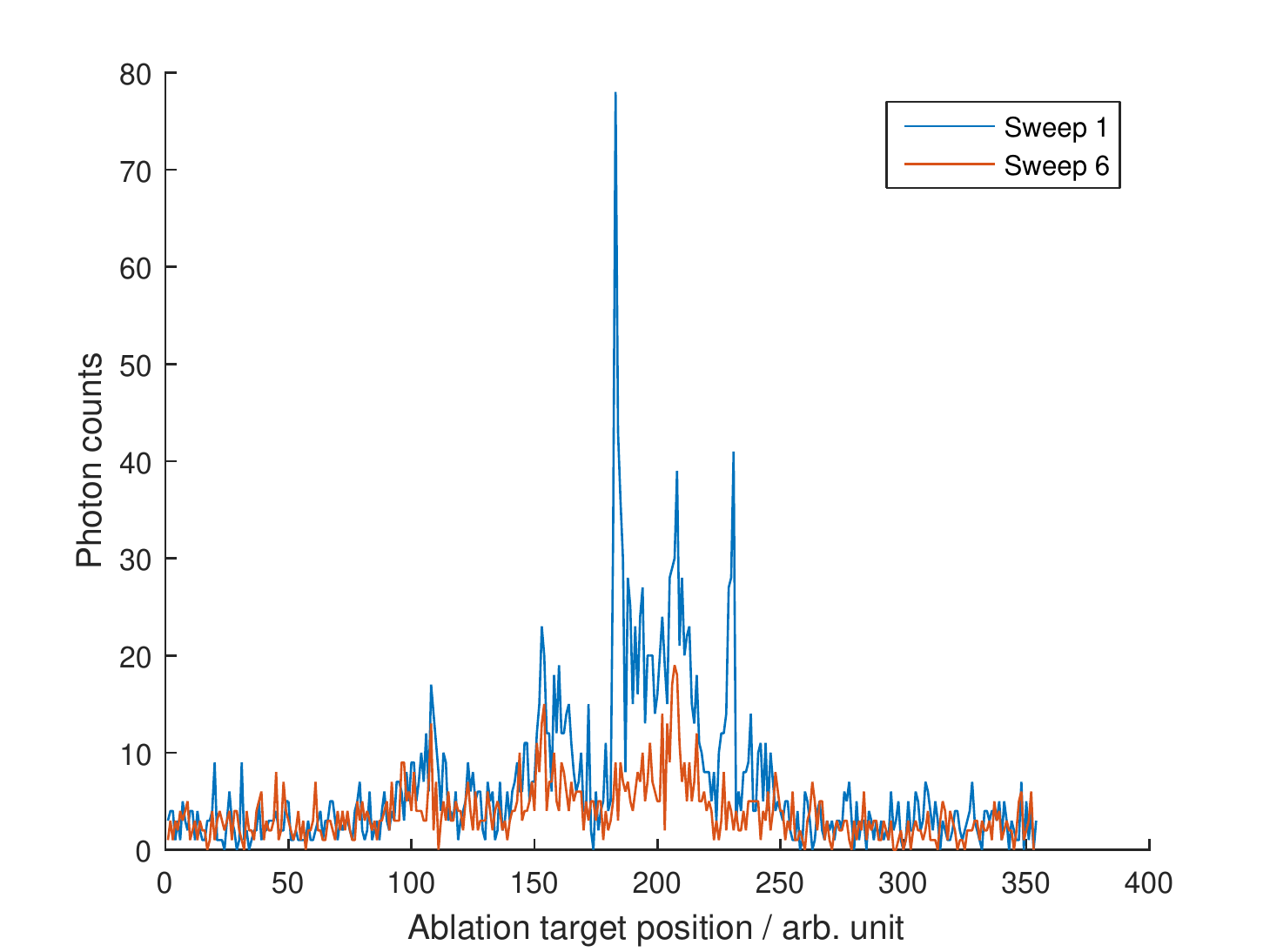}}
	\subfloat[\label{sfig:RadNeutralFluorescence_207uJ}]{\includegraphics[
		width=0.4\linewidth]{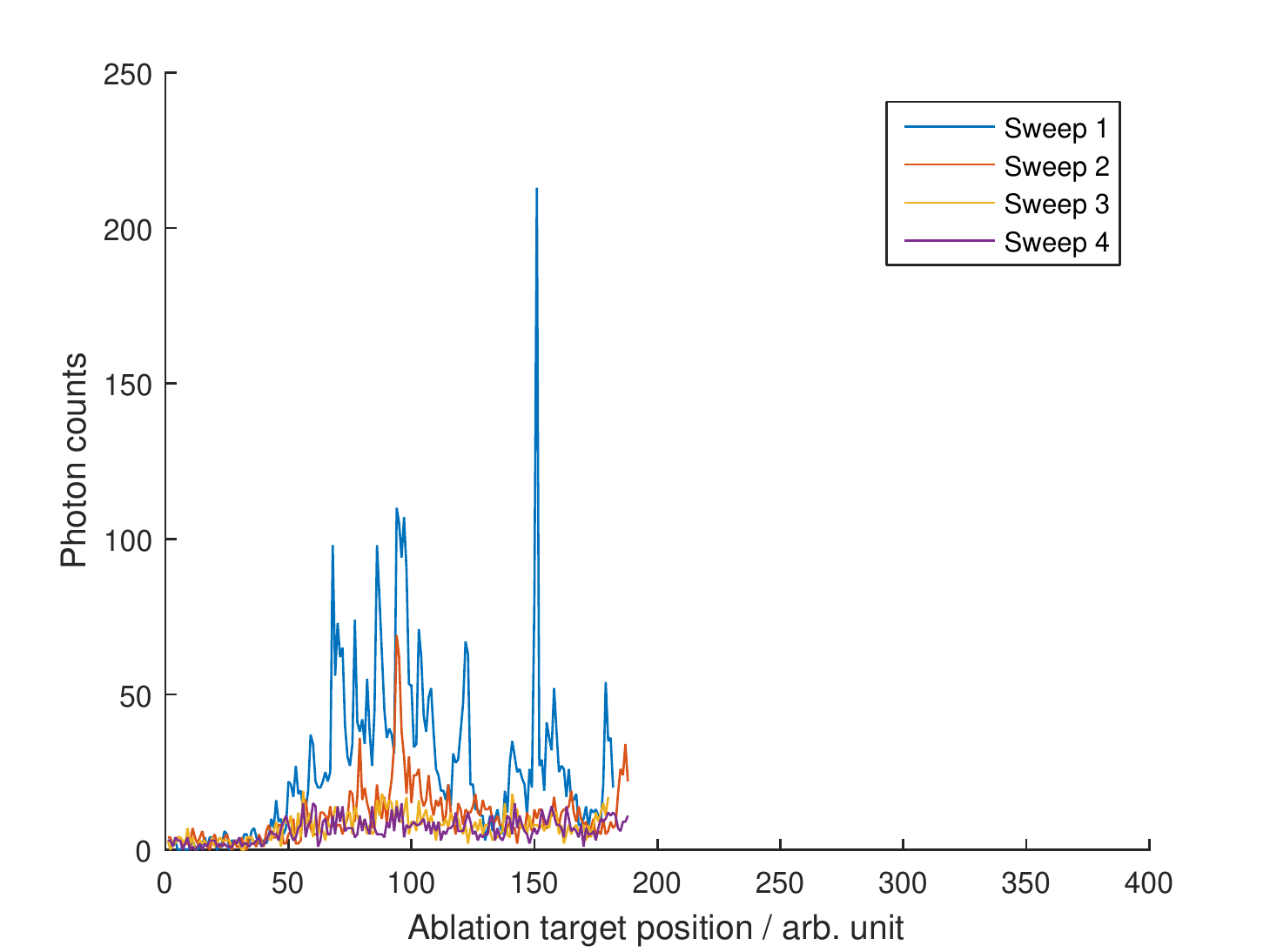}}
	\hfill
	\subfloat[\label{sfig:RadNeutralFluorescence_305uJ}]{\includegraphics[
		width=0.4\linewidth]{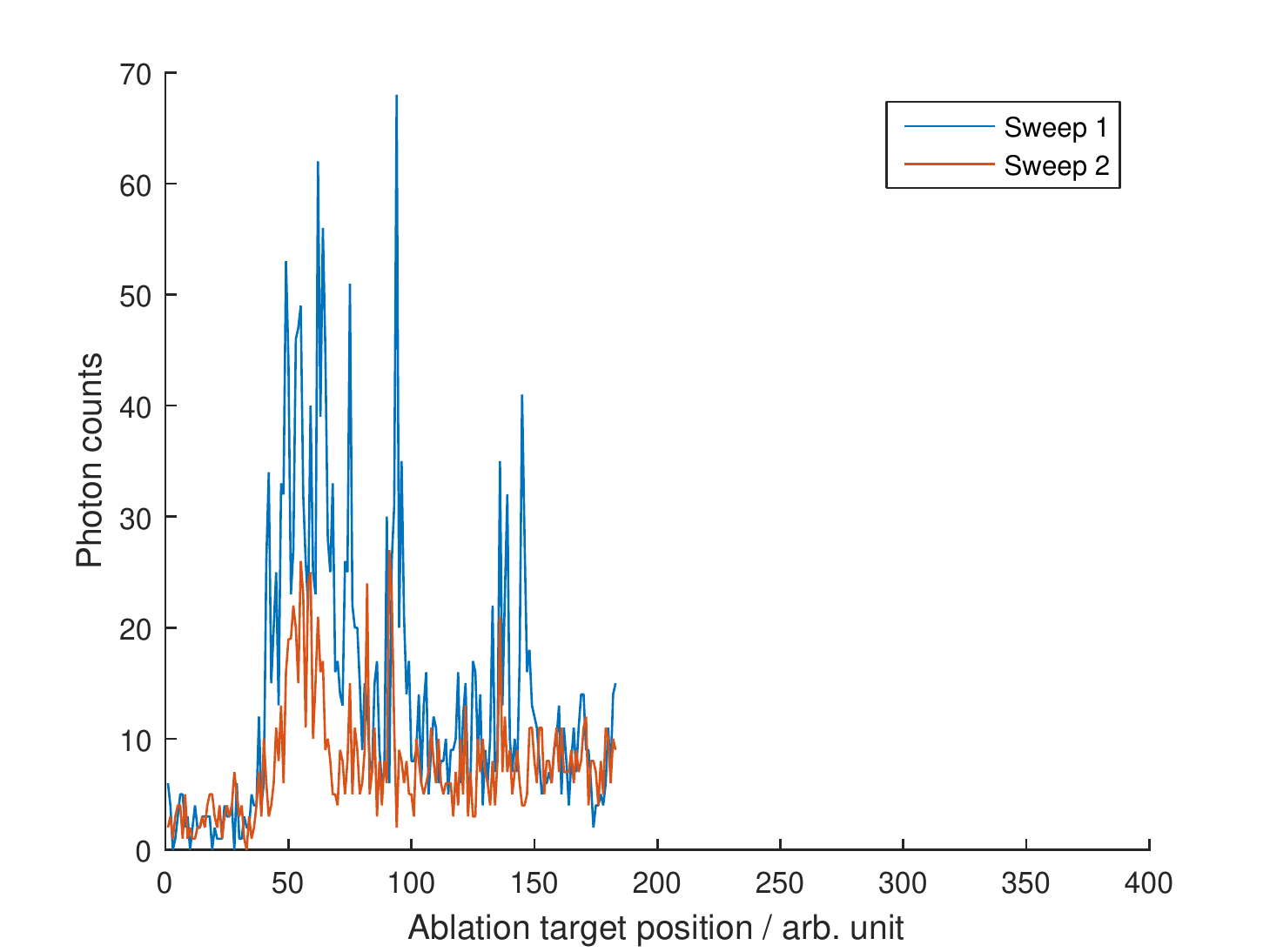}}
	\subfloat[\label{sfig:RadNeutralFluorescence_125uJ_afterhighpower}]{\includegraphics[
		width=0.4\linewidth]{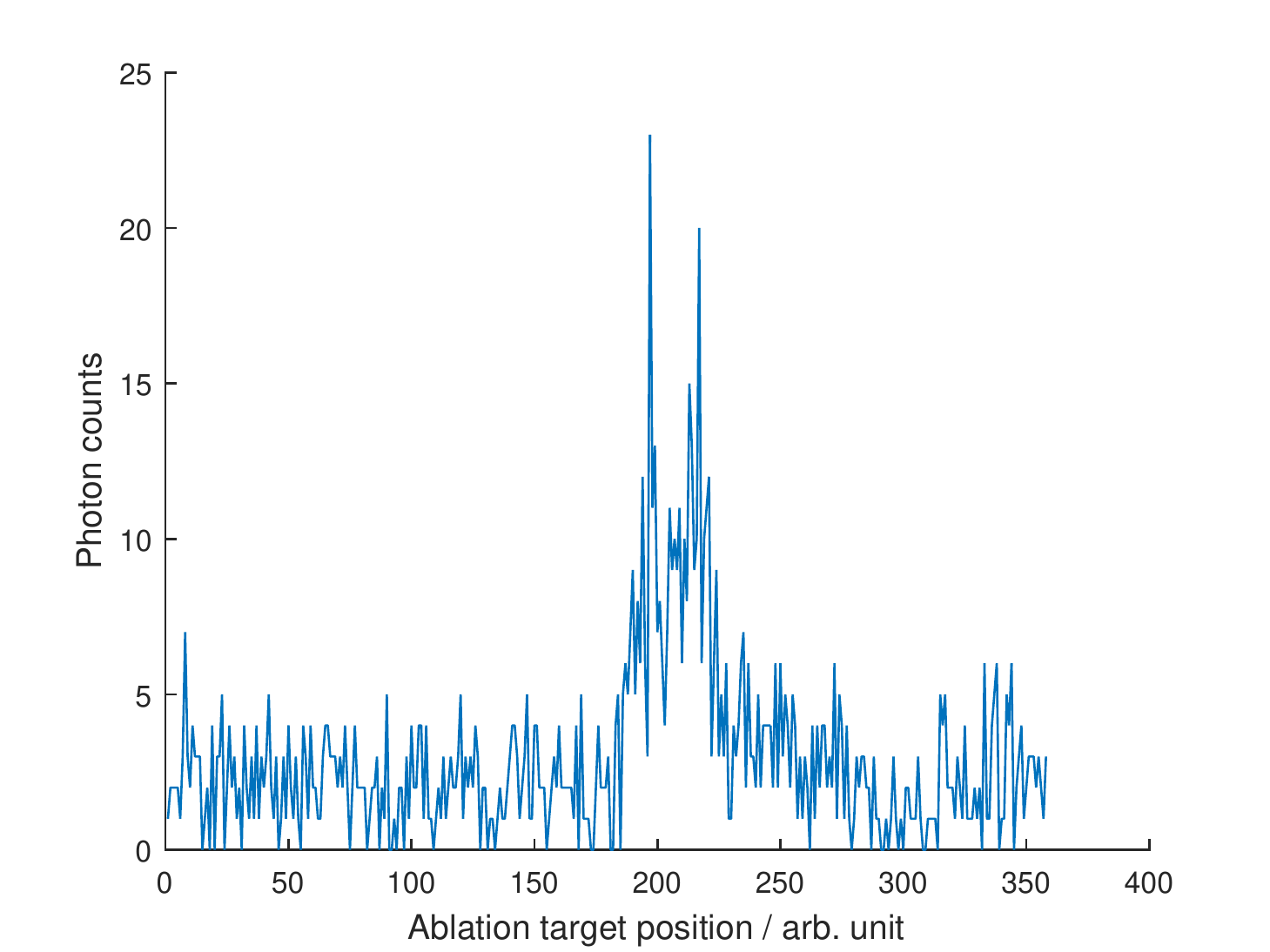}}
	\caption{
		Neutral barium fluorescence from the low-volume target, sweeping the ablation laser along a line on the target. 
		(a) Initial full-line sweeps at ablation pulse fluence of $\SI{0.59}{\joule\per\centi\meter\squared}$. 
		(b) Half-line sweeps at ablation pulse fluence of $\SI{0.98}{\joule\per\centi\meter\squared}$. 
		(c) Half-line sweeps at ablation pulse fluence of $\SI{1.44}{\joule\per\centi\meter\squared}$. 
		(d) A full-line sweep at ablation pulse fluence of $\SI{0.59}{\joule\per\centi\meter\squared}$ after the half-line sweeps in (c) and (d).}
	\label{fig:RadNeutralFluorescence}
\end{figure*}

On the first sweep of the ablation laser along the line on the target, the peak fluorescence count detected is approximately 80. 
On repeat sweep of the ablation laser along the same line, the fluorescence counts decrease rapidly. 
By the sixth sweep, the peak fluorescence count has dropped to below 20. 
In an effort to try to condition the target, similar to the procedure done for the high-volume target, the pulse fluence is increased to $\SI{0.98}{\joule\per\centi\meter\squared}$ and the ablation laser is swept for half of the line. 
The fluorescence counts are high for the initial scan, at approximately 200 peak fluorescence count, but quickly dropped to below 20 counts on the fourth sweep. 
This is contrary to the high-volume target, where the neutral barium fluorescence counts would gradually increase and stabilize at a consistent count during the conditioning process. 
The pulse fluence is further increased to $\SI{1.44}{\joule\per\centi\meter\squared}$ for the half-line sweep. 
The peak fluorescence count is approximately 70 for the first can, but quickly drops to below 30 on the second scan. 
Going back to the lower pulse fluence at $\SI{0.59}{\joule\per\centi\meter\squared}$ and performing the full-line scan, we observe that the half-line that went through ablations at high pulse fluence no longer produce any neutral barium fluorescence. 
In contrast to the high-volume target, instead of conditioning a spot, it shows signs of a spot being depleted of the source material when high pulse fluence is used to ablate it. 
Attempts to trap $^{138}\mathrm{Ba}^+$ (which has an abundance of 53.3\% on the low-volume target) using the two-step photoionization method follow the same trend, where trapping quickly becomes impossible in the order of tens of pulses, indicating that the source material is depleted.

\begin{figure}[H]
	\centering
	\includegraphics[width=\linewidth]{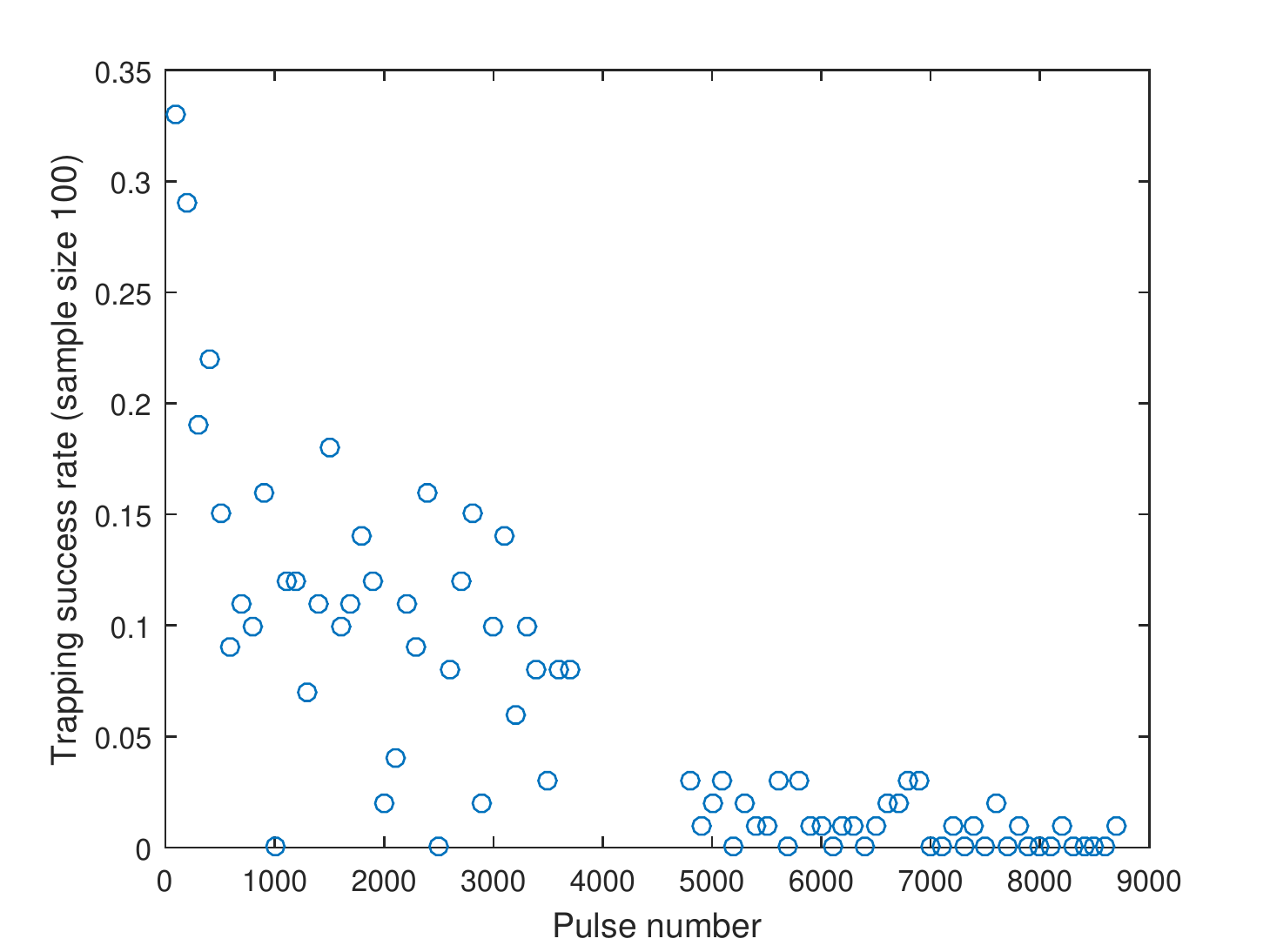}
	\caption{
		Trapping probability of $^{138}\mathrm{Ba}^+$ from the low-volume target using the direct-ion loading method. 
		The horizontal axis denotes the pulse number that is sent to the same spot on the target. 
		The vertical axis shows the trapping success rate out of a sample size of 100. 
		The ablation fluence used is $\SI{3.84}{\joule\per\centi\meter\squared}$.}
	\label{fig:Rad_Direct-ion_trapping_prob}
\end{figure}

To test the low-volume target further, we employ the trapping procedure used by Reference \cite{Hucul2017}. 
The ablation laser fluence is increased to $\SI{3.84}{\joule\per\centi\meter^2}$ to directly generate barium ions. 
Using the direct ion-loading method, the ion trapping success rate of $^{138}\mathrm{Ba}^+$ is initially in the order of 10\% initially (see Figure \ref{fig:Rad_Direct-ion_trapping_prob}).
However, it quickly declines to the order of 1\% after a few thousand pulses. 
After about 10,000 pulses, the trapping probability is below 1\%. 
This indicates that a target spot is being depleted quickly with this trapping procedure as well. 
Furthermore, we are only able to trap single ions and not ion chains as described in \cite{Hucul2017}.
Therefore, with this low-volume target and both loading methods, it is impossible to establish an ion chain.
Given the poor trapping performance, low spot lifetimes, and overall expected target lifetime, this target is impractical for loading ions for quantum computing experiments.

\section{Observation of other excited-state neutral species} \label{sec:Other_excited_state_species}
By ablating the barium targets at high pulse fluences, on the order of $\SI{1}{\joule\per\centi\meter^2}$, we also note the presence of some fluorescence after ablation that is not driven by the $\SI{554}{\nano\meter}$ nor the $\SI{493}{\nano\meter}$ lasers, indicating that they are not barium atoms/ions. 
In these experiments, ablation pulses of pulse fluence $\SI{3.84}{\joule\per\centi\meter^2}$ are sent to ablate the low-volume target. 
An imaging system collects light from the trap center. 
The collected light is detected using a PMT. 
The PMT is set to detect photons for a time window of $\SI{1}{\micro\second}$, starting at different times relative to the arrival time of the ablation pulse to do time-resolved light detection. 
Different spectral filters are installed in the imaging system to investigate the origin of the detected light. 
The trap voltages are turned off to allow both charged and uncharged particles to go through the trap.

\begin{figure*}[h]
	\centering
	\subfloat[\label{sfig:TOF_rad_1x_vs_2x_532filter}]{\includegraphics[width=0.45\linewidth]{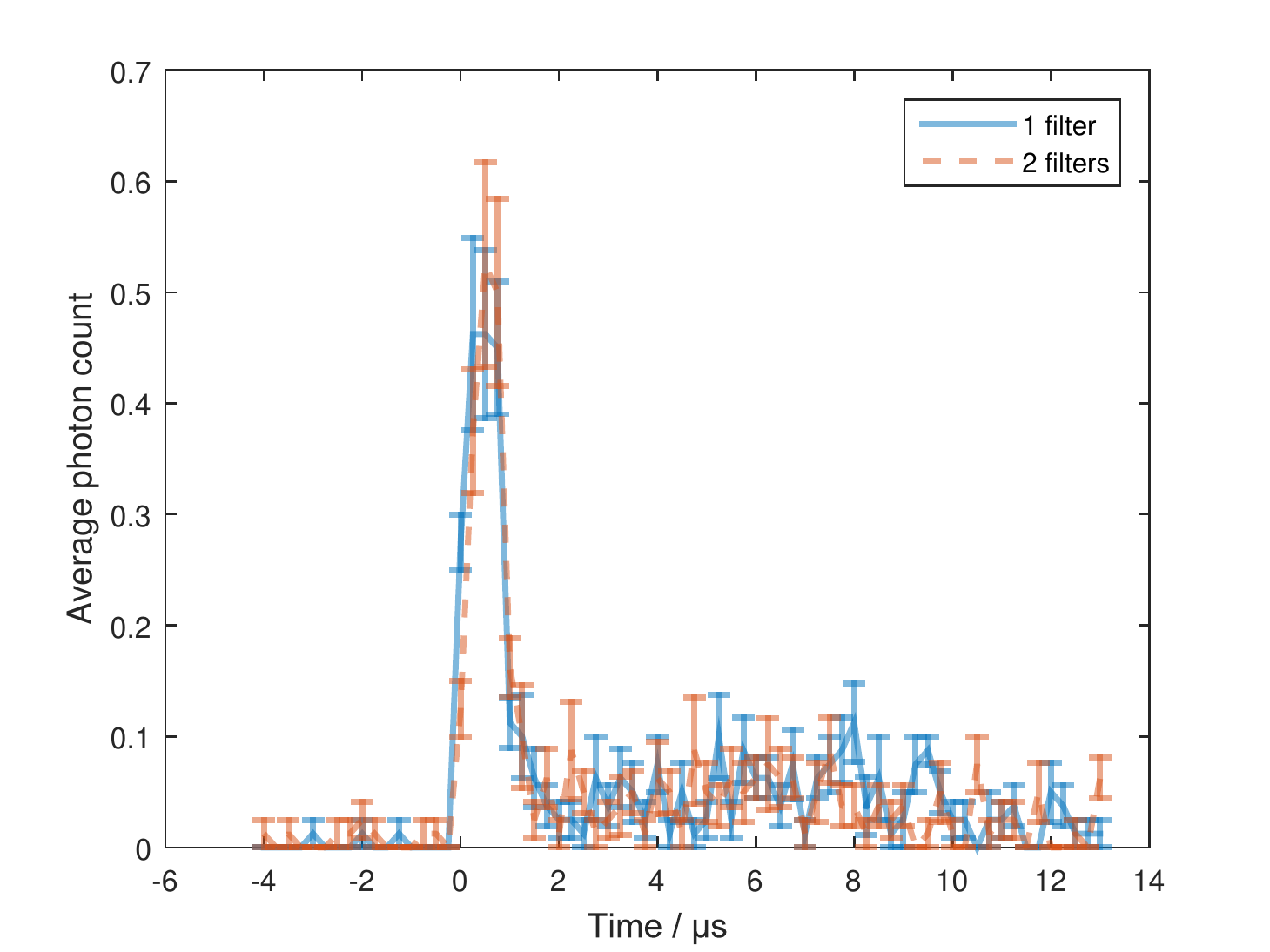}}
	\subfloat[\label{sfig:TOF_rad_no_vs_with_550LPfilter}]{\includegraphics[width=0.45\linewidth]{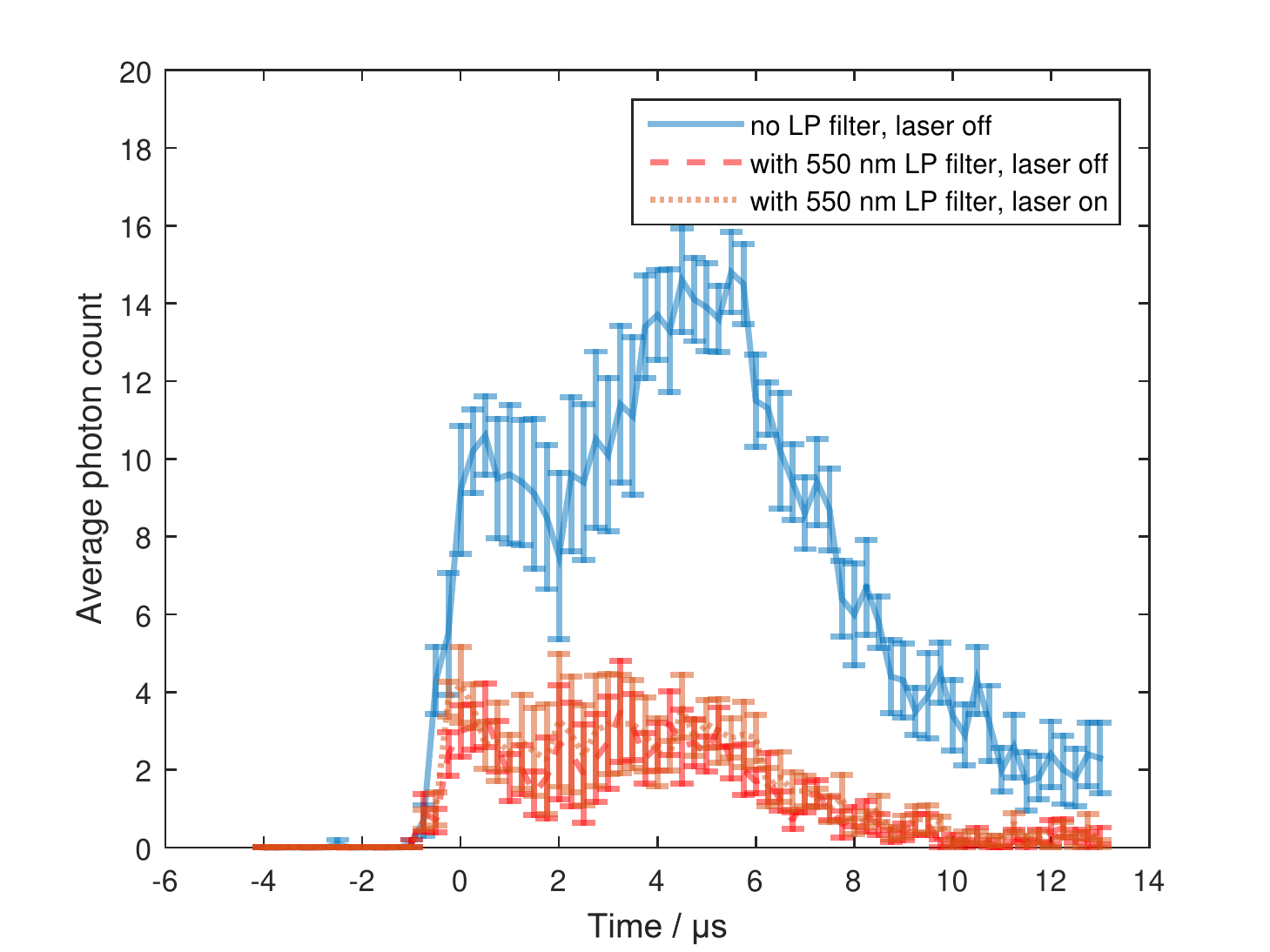}}
	\hfill
	\subfloat[\label{sfig:TOF_rad_no_vs_with_493_650_lasers}]{\includegraphics[width=0.45\linewidth]{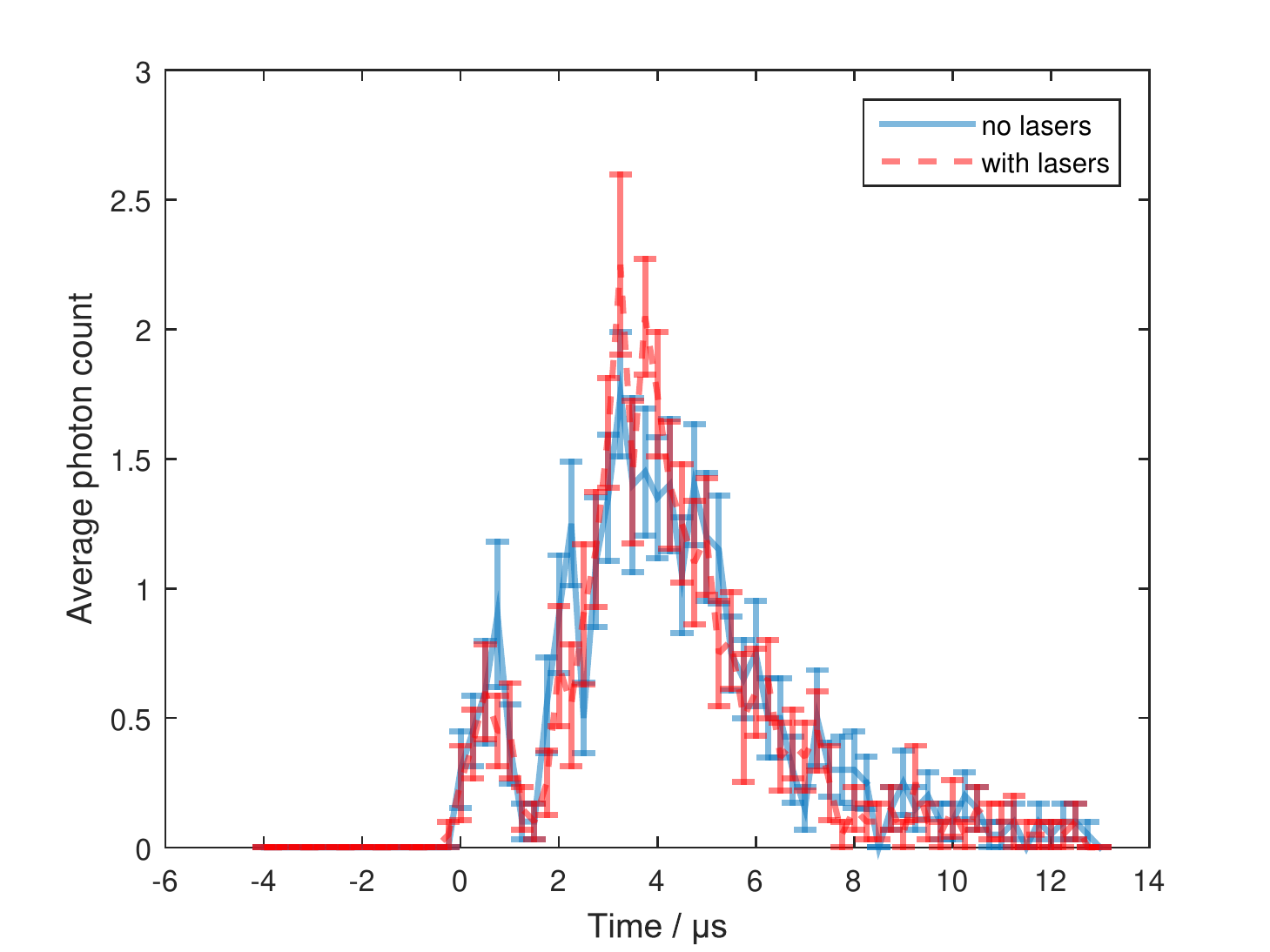}}
	\subfloat[\label{sfig:TOF_rad_no_vs_with_trap_voltage}]{\includegraphics[width=0.45\linewidth]{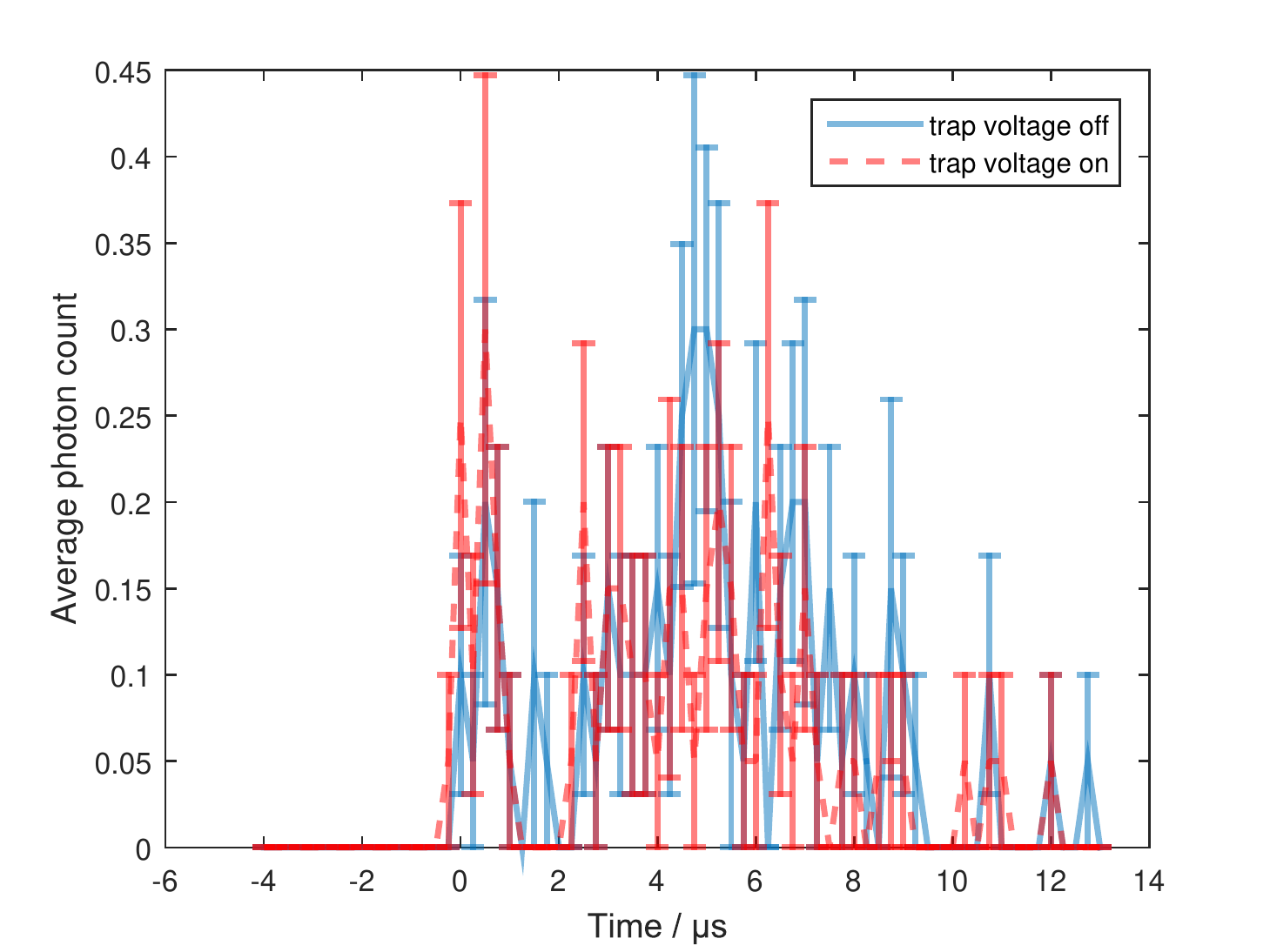}}
	\caption{
		Time-resolved light detection after ablating the low-volume target with a pulse fluence of $\SI{3.84}{\joule\per\centi\meter^2}$. 
		(a) Comparison of signals between having 1 and 2 $\SI{532}{\nano\meter}$ notch filter in the imaging system.
		No qualitatively significant difference is found. 
		(b) Comparison of signals between having and not having a $\SI{550}{\nano\meter}$ longpass filter in the imaging system as well as with and without the $\SI{554}{\nano\meter}$ laser turned on. 
		The signal is partially attenuated with the $\SI{550}{\nano\meter}$ longpass filter. 
		No practical difference is observed with or without the $\SI{553.7}{\nano\meter}$ laser turned on. 
		(c) Comparison of signals with and without the barium ion fluorescence lasers turned on. 
		A $\SI{488}{\nano\meter}$ bandpass filter is installed in the imaging system. 
		No qualitatively significant difference is found. 
		(d) Comparision of signals with and without the trap voltages turned on. 
		A $\SI{488}{\nano\meter}$ bandpass filter is installed in the imaging system. 
		No qualitatively significant difference is found.}
	\label{fig:Rad_target_excited_fluorescence}
\end{figure*}
\begin{figure}[h]
	\centering
	\subfloat[\label{sfig:TOF_nat_with_vs_no_553laser80uWCB_with_vs_no_550LP}]{\includegraphics[width=\linewidth]{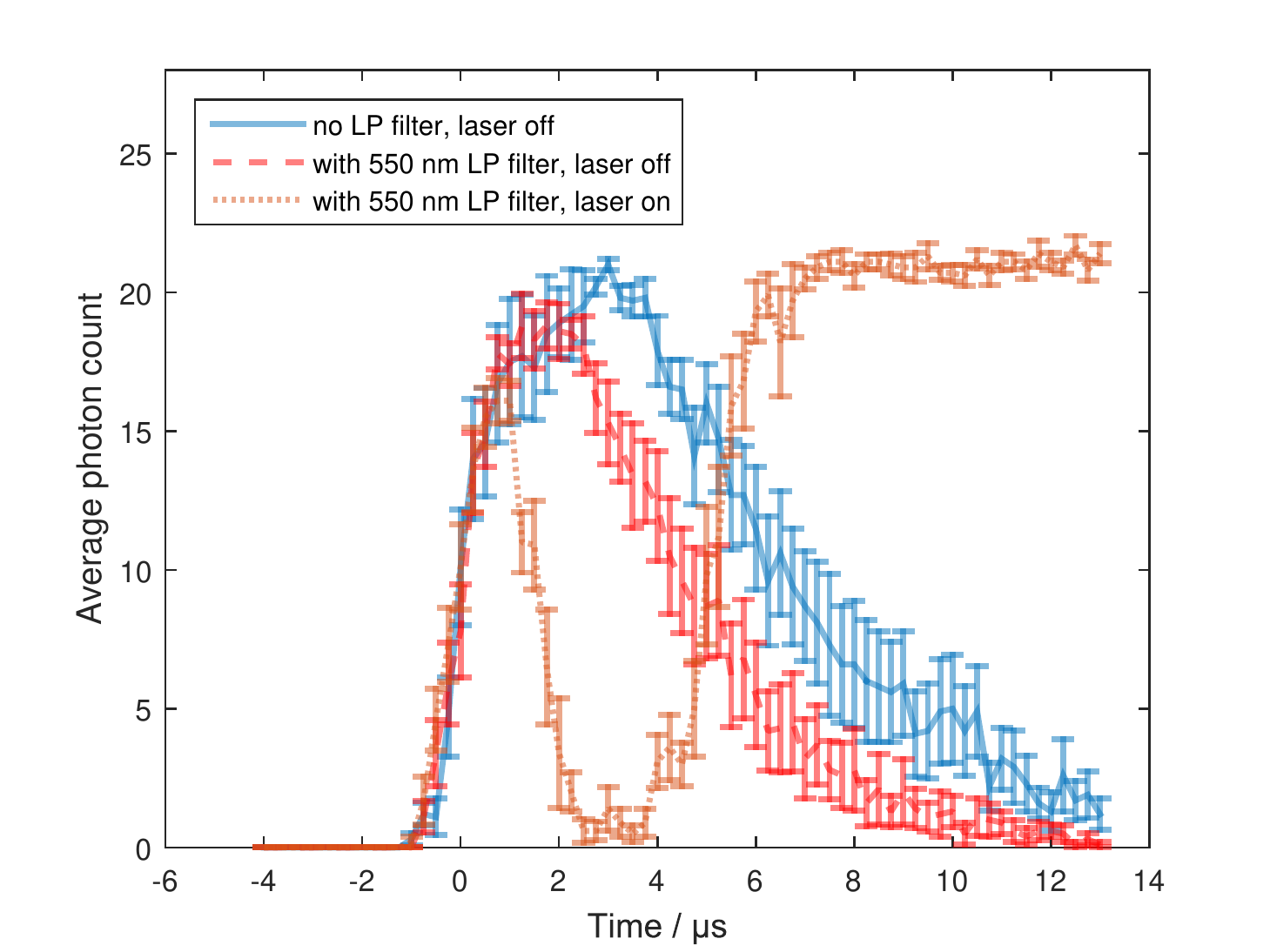}}
	\hfill
	\subfloat[\label{sfig:Nat_target_hole}]{\includegraphics[width=0.6\linewidth]{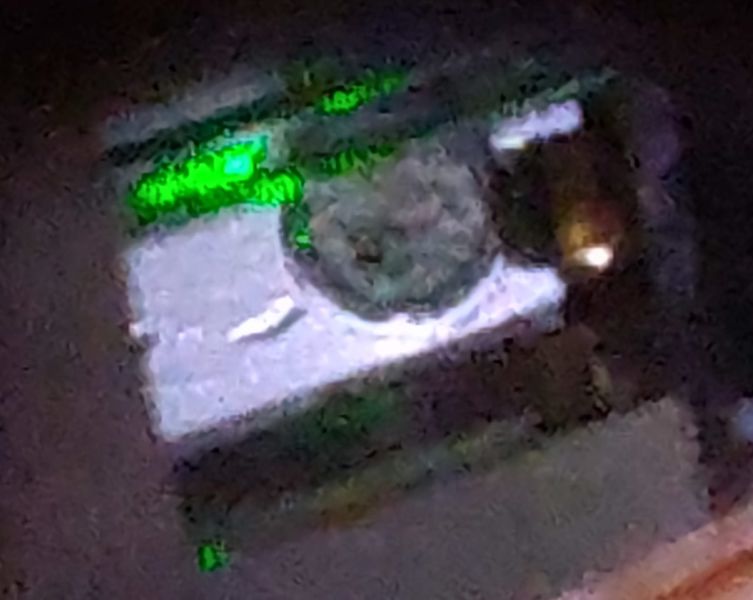}}
	\caption{
		(a) Time-resolved light detection after ablating the high-volume target with a pulse fluence of $\SI{3.84}{\joule\per\centi\meter^2}$. 
		Comparison of signals between having and not having a $\SI{550}{\nano\meter}$ longpass filter in the imaging system as well as with and without the $\SI{554}{\nano\meter}$ laser turned on. 
		The signal is partially attenuated with the $\SI{550}{\nano\meter}$ longpass filter. 
		With the $\SI{554}{\nano\meter}$ laser turned on, the early time signal is attenuated while the late time signal is enhanced. 
		(b) Charred spot on the high-volume target after ablating with a pulse fluence of $\SI{3.84}{\joule\per\centi\meter^2}$.}
	\label{fig:Nat_target_excited_fluorescence}
\end{figure}
We first observe that there is a photon count peak right after the ablation laser reaches the target, even when all the CW lasers are turned off. 
We verify that this peak is not $\SI{532}{\nano\meter}$ light by observing that the photon count does not change significantly whether there are 1 or 2 $\SI{532}{\nano\meter}$ notch filters (Thorlabs NF533-17) in the imaging system. 
We also confirm that it is not residual $\SI{1064}{\nano\meter}$ light from the ablation laser by observing that the peak disappears when a $\SI{1000}{\nano\meter}$ longpass filter (Thorlabs FEL1000) is introduced into the imaging system. 
Our conjecture is that this could be due some up- or down-conversion of the high energy $\SI{532}{\nano\meter}$ pulse, shifting it from the $\SI{532}{\nano\meter}$ frequency and thus getting pass the $\SI{532}{\nano\meter}$ notch filter.

An interesting observation is that there are photon counts detected that last several tens of microseconds after the arrival of the ablation pulse when no CW lasers are turned on.
All the experiments discussed from here on in this section are done with $\SI{532}{\nano\meter}$ notch filter to filter off light from the ablation laser. 
First, we observe that the signal is partially reduced upon the addition of a $\SI{550}{\nano\meter}$ longpass filter (Thorlabs FEL0550). 
To see if the fluorescence is from barium ions, the experiments are repeated with and without the $\SI{493}{\nano\meter}$ and $\SI{650}{\nano\meter}$ turned on. 
The imaging system is installed with a $\SI{488}{\nano\meter}$ narrow bandpass filter (Brightline FF01-488/10-25), which allows wavelengths in the range from $\SI{481}{\nano\meter}$ to $\SI{495}{\nano\meter}$ to pass through. 
It is observed that the signals are not qualitatively different with or without the lasers turned on, indicating that they are likely not barium ions. 
To further verify if they are ions, the experiments are repeated with and without the trap voltages being turned on.
The same $\SI{488}{\nano\meter}$ narrow bandpass filter is installed and the CW lasers are turned off. 
No significant qualitative difference in the signals observed is found, whether the trap voltages are turned on or not. 
This indicates that the fluorescing particles are uncharged. 
We thus conclude that they are uncharged particles in some excited states, releasing photons as they pass through the trap.

We also repeated the time-resolved experiments are also performed on the high-volume target. 
Three sets of experiments are done, with the following combinations: without $\SI{550}{\nano\meter}$ longpass filter and without $\SI{554}{\nano\meter}$ laser, with $\SI{550}{\nano\meter}$ longpass filter and without $\SI{554}{\nano\meter}$ laser, and with $\SI{550}{\nano\meter}$ longpass filter and with $\SI{554}{\nano\meter}$ laser. 
Similar to the low-volume target, the signal observed is partially reduced with the $\SI{550}{\nano\meter}$ longpass filter on, without the $\SI{554}{\nano\meter}$ laser. 
With the $\SI{554}{\nano\meter}$ laser turned on, the signals are earlier times (before $\SI{6}{\micro\second}$) are reduced and the signals at later times (after $\SI{6}{\micro\second}$) are increased. 
Our conjecture is that the signals at earlier times are from excited state species. 
With the $\SI{554}{\nano\meter}$ laser turned on, they are further excited or ionized, leading to them fluorescing at a lower wavelength below $\SI{550}{\nano\meter}$, or are in an ionic ground state. 
In both cases, the signal at the PMT would be reduced. 
The increased signal at later times are most likely coming from neutral barium ions. 
We also find that it is not advisable to ablate the high-volume target at this high pulse fluence of $\SI{3.84}{\joule\per\centi\meter^2}$, as it left a visible char mark after these sets of experiments. 
This is also the reason why no further time-resolved light detection experiments at this pulse fluence is done on the high-volume target.

\section{$^{137}\mathrm{Ba}^+$ selectivity} \label{sec:Ba-137 Selectivity}
In order to characterize isotope selectivity for $^{137}\mathrm{Ba}^+$, ion chains are loaded into the trap and the number of $^{137}\mathrm{Ba}^+$ compared to the dark ions are counted. 
The experimental procedure is as follows. 
The barium atom first excitation ($\SI{554}{\nano\meter}$) laser frequency is blue-detuned by $\SI{275}{\mega\hertz}$ from the neutral barium fluorescence peak of $^{138}\mathrm{Ba}^+$. 
This $\SI{554}{\nano \meter}$ laser frequency corresponds to one of the $^{137}\mathrm{Ba}^+$ isotope fluorescence peaks. 
The cooling ($\SI{493}{\nano\meter}$) laser frequencies are set to drive the transitions $\ket{S,F=1} \leftrightarrow \ket{P,F=2}$ and $\ket{S,F=2} \leftrightarrow \ket{P,F=2}$.
This is done by blue-detuning the $\SI{493}{\nano\meter}$ laser carrier frequency by $\SI{1830}{\mega\hertz}$ from the optimal frequency for $^{138}\mathrm{Ba}^+$ ion fluorescence and generating the sidebands required for the $^{137}\mathrm{Ba}^+$ transitions using an EOM at the modulation frequency of $\SI{4019}{\mega\hertz}$. 
The repump ($\SI{650}{\nano\meter}$) laser frequencies are set to drive the transitions $\ket{D,F=0} \leftrightarrow \ket{P,F=1}$, $\ket{D,F=1} \leftrightarrow \ket{P,F=1}$, $\ket{D,F=2} \leftrightarrow \ket{P,F=1}$ and $\ket{D,F=3} \leftrightarrow \ket{P,F=2}$. 
This is done by red-detuning the carrier frequency of the $\SI{650}{\nano\meter}$ laser by $\SI{770}{\mega\hertz}$ from the optimal frequency for $^{138}\mathrm{Ba}^+$ ion fluorescence and generating the sidebands with an EOM at frequency modulations of $\SI{314}{\mega\hertz}$, $\SI{470}{\mega\hertz}$ and $\SI{874}{\mega\hertz}$. 
This transition scheme is chosen so that all the extra sideband frequencies are red-detuned from the $\ket{D} \leftrightarrow \ket{P}$ transitions to prevent ion heating. 
The first excitation, cooling, repump and an ionizing laser ($\SI{405}{\nano\meter}$) are sent to the center of our four-rod trap. 
The trap voltages are turned on for ion trapping. Ablation pulses at a pulse fluence of $\SI{0.25}{\joule\per\centi\meter\squared}$ are then sent to the barium salt target. 
For each trapping attempt, 10 ablation pulses at a repetition rate of $\SI{2}{\hertz}$ is used. 
After each attempt, the number of bright ions trapped are checked with a CMOS camera. 
The trapping attempts are repeated until 3 bright ions are observed. 
When 3 bright ions are trapped, the repump laser is toggled off and on for 100 times, and the ion image on the CMOS camera is captured after each toggling. 
This helps to shift the positions of the ions in the ion chain, so that we can capture all the ion positions in the chain with the trapped bright ions.

\begin{figure*}
	\centering
	\subfloat[\label{sfig:Ba137_Sum_image_01}]{\includegraphics[width=0.25\linewidth]{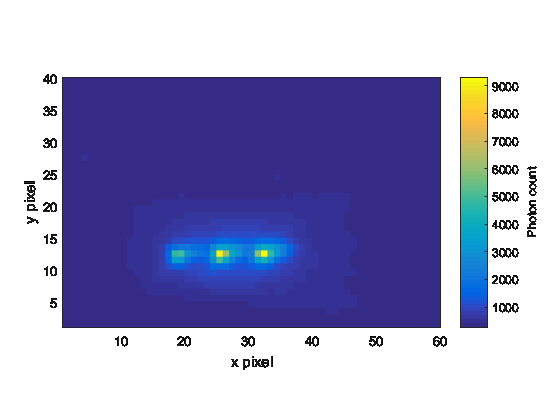}}
	\subfloat[\label{sfig:Ba137_Sum_image_02}]{\includegraphics[width=0.25\linewidth]{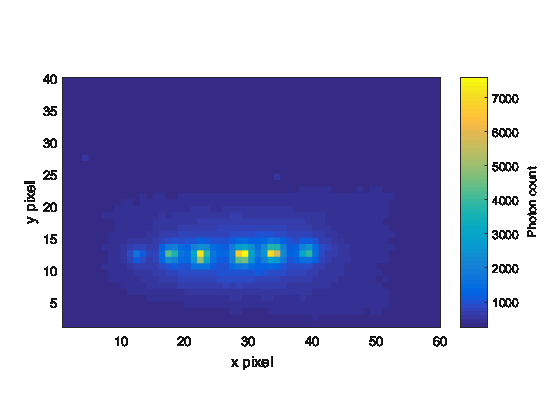}}
	\subfloat[\label{sfig:Ba137_Sum_image_03}]{\includegraphics[width=0.25\linewidth]{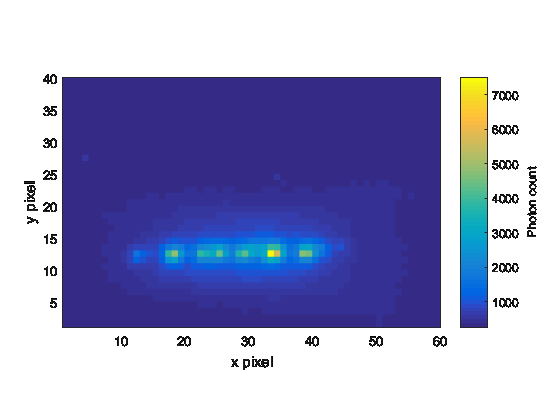}}
	\subfloat[\label{sfig:Ba137_Sum_image_04}]{\includegraphics[width=0.25\linewidth]{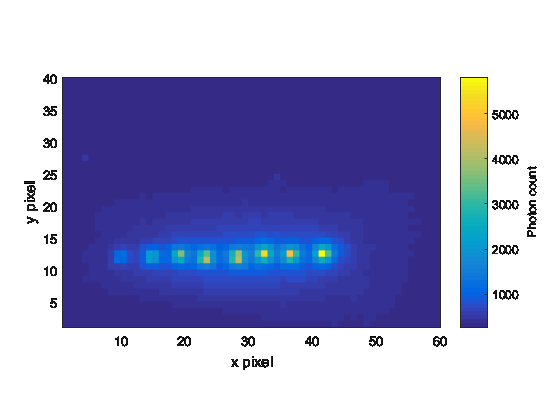}}
	\hfill
	\subfloat[\label{sfig:Ba137_Sum_image_05}]{\includegraphics[width=0.25\linewidth]{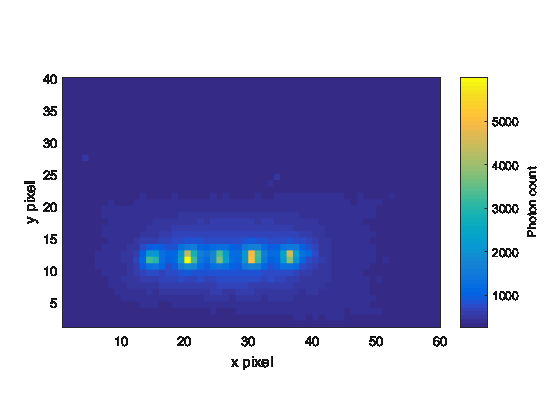}}
	\subfloat[\label{sfig:Ba137_Sum_image_06}]{\includegraphics[width=0.25\linewidth]{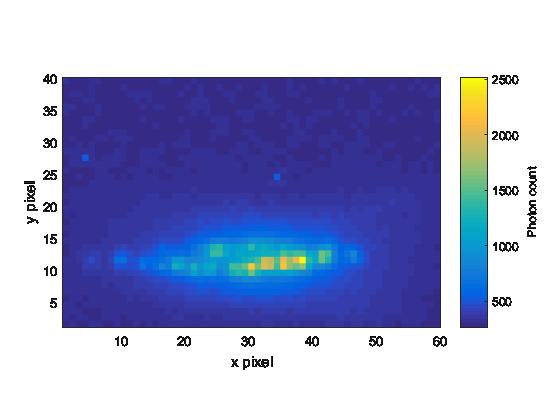}}
	\subfloat[\label{sfig:Ba137_Sum_image_07}]{\includegraphics[width=0.25\linewidth]{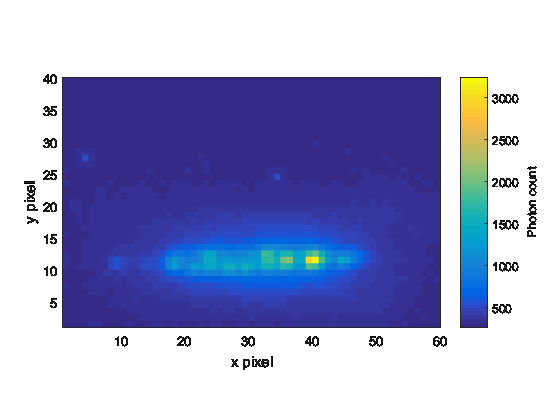}}
	\subfloat[\label{sfig:Ba137_Sum_image_08}]{\includegraphics[width=0.25\linewidth]{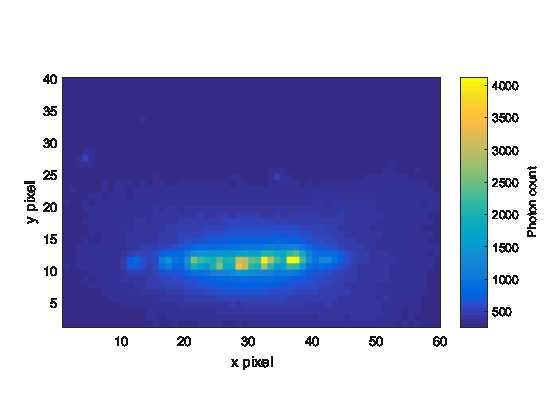}}
	\hfill
	\subfloat[\label{sfig:Ba137_Sum_image_09}]{\includegraphics[width=0.25\linewidth]{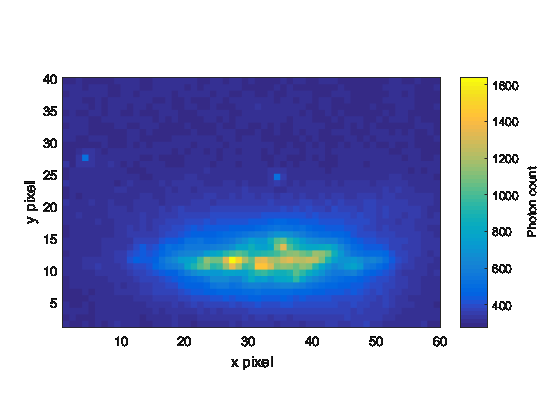}}
	\subfloat[\label{sfig:Ba137_Sum_image_10}]{\includegraphics[width=0.25\linewidth]{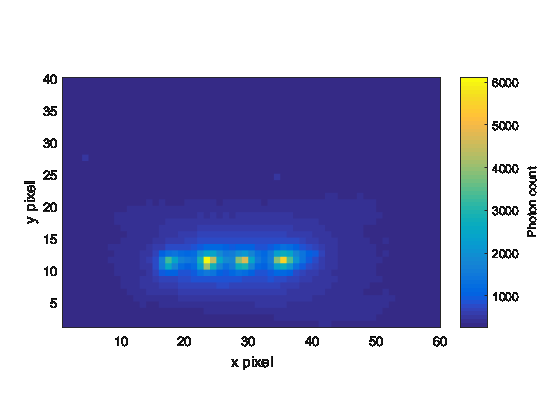}}
	\caption{Ion chain images from summing across all images from toggling the repump laser. Each image is a separate experiment. The first excitation laser ($\SI{554}{\nano\meter}$ laser) intensity is set to $\SI{0.104}{\watt\per\centi\meter\squared}$. The total number of ions (bright and dark) are manually counted to be (a) 3, (b) 7, (c) 7, (d) 8, (e) 5, (f) 14, (g) 12, (h) 8, (i) 13, and (j) 4.}
	\label{fig:Ba137_Sum_images_4uW}
\end{figure*}
To count the total number of ions in an ion chain, the 100 images from toggling the repump laser are summed together to form a single image which shows all the positions the bright ions has hopped to during the repump laser toggling. 
The number of ion positions are then counted manually. 
The ion-loading selectivity is then estimated by dividing the number of bright ions trapped with the number of ion positions detected.

The experiments are repeated 10 times with the first excitation laser at an intensity of $\SI{0.104}{\watt\per\centi\meter\squared}$. 
The aggregated images to show the ion positions are presented in Figure \ref{fig:Ba137_Sum_images_4uW}. 
Across all 10 ion chains, a total of 30 bright ions are detected and the total number of bright and dark ions are counted to be 81. 
This gives a selectivity estimate of $37 \pm 5 \%$ for $^{137}\mathrm{Ba}^+$.

\begin{figure*}
	\centering
	\subfloat[\label{sfig:Ba137_Sum_image_36uW_01}]{\includegraphics[width=0.25\linewidth]{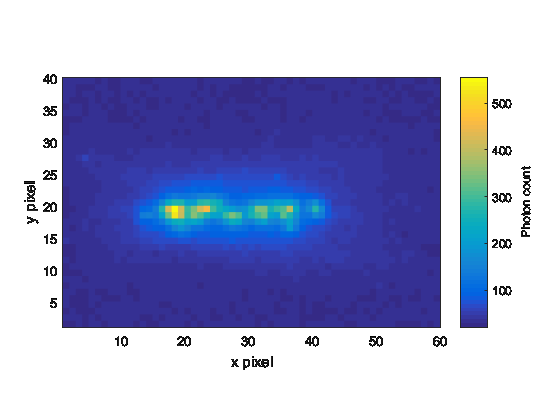}}
	\subfloat[\label{sfig:Ba137_Sum_image_36uW_02}]{\includegraphics[width=0.25\linewidth]{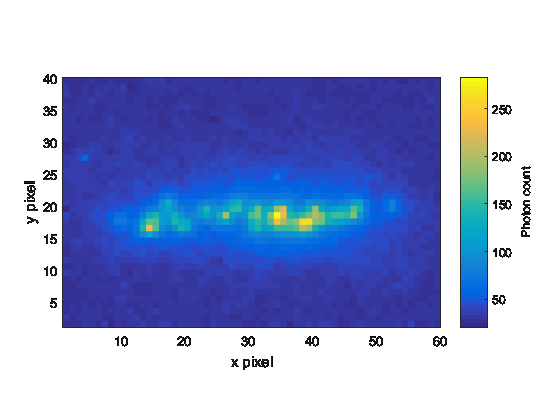}}
	\subfloat[\label{sfig:Ba137_Sum_image_36uW_03}]{\includegraphics[width=0.25\linewidth]{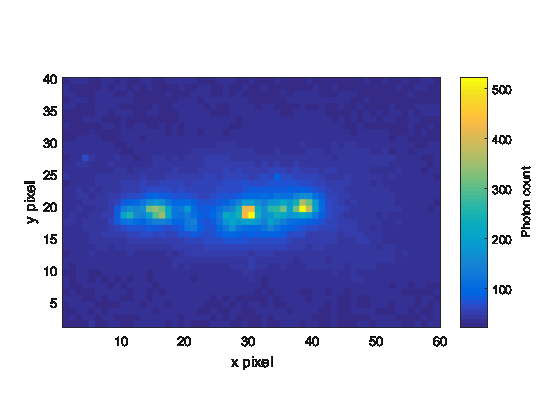}}
	\subfloat[\label{sfig:Ba137_Sum_image_36uW_04}]{\includegraphics[width=0.25\linewidth]{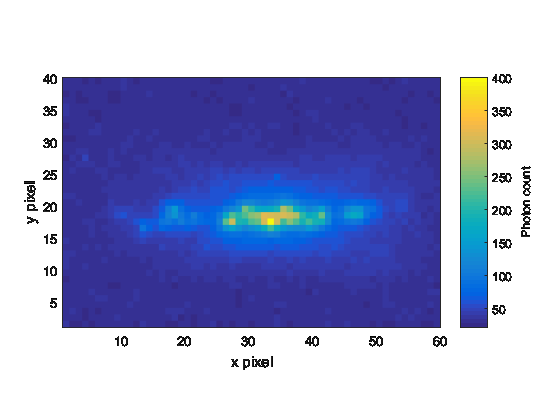}}
	\hfill
	\subfloat[\label{sfig:Ba137_Sum_image_36uW_05}]{\includegraphics[width=0.25\linewidth]{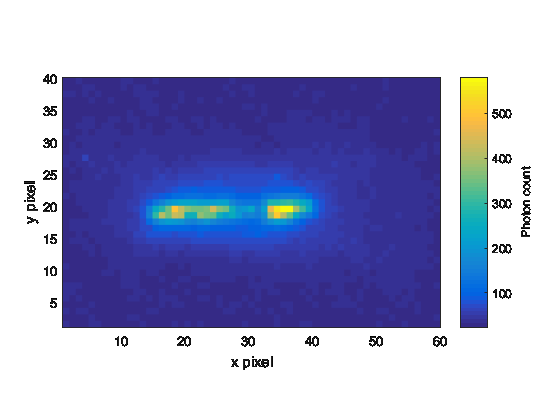}}
	\subfloat[\label{sfig:Ba137_Sum_image_36uW_06}]{\includegraphics[width=0.25\linewidth]{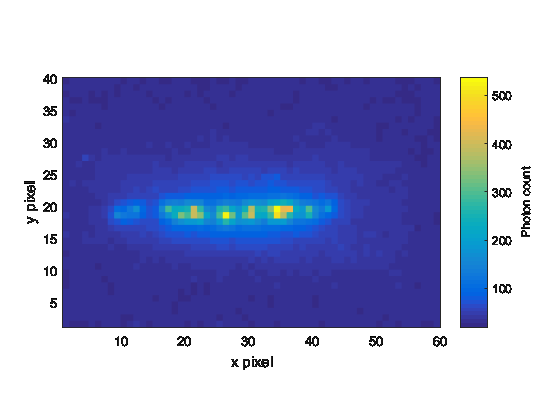}}
	\subfloat[\label{sfig:Ba137_Sum_image_36uW_07}]{\includegraphics[width=0.25\linewidth]{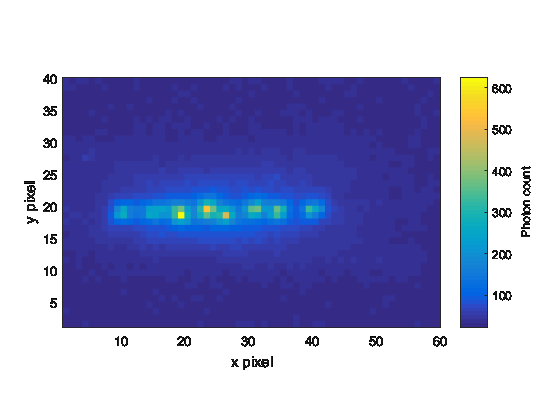}}
	\subfloat[\label{sfig:Ba137_Sum_image_36uW_08}]{\includegraphics[width=0.25\linewidth]{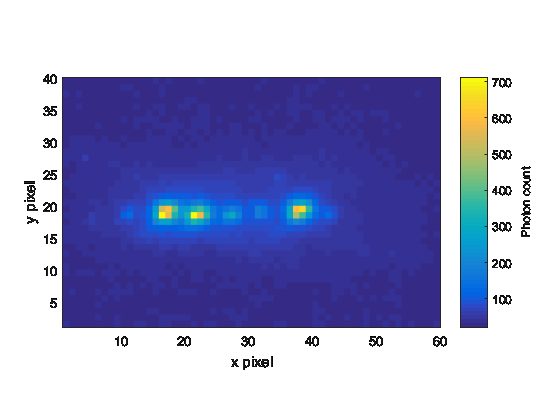}}
	\hfill
	\subfloat[\label{sfig:Ba137_Sum_image_36uW_09}]{\includegraphics[width=0.25\linewidth]{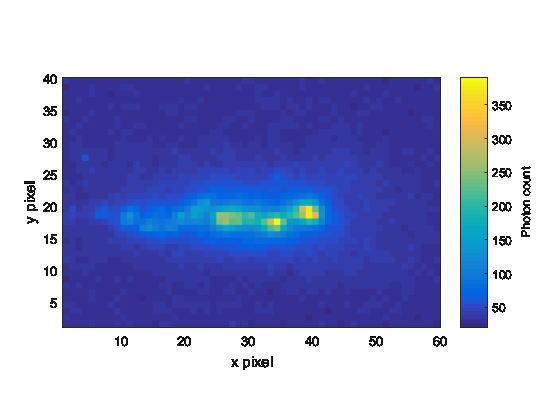}}
	\subfloat[\label{sfig:Ba137_Sum_image_36uW_10}]{\includegraphics[width=0.25\linewidth]{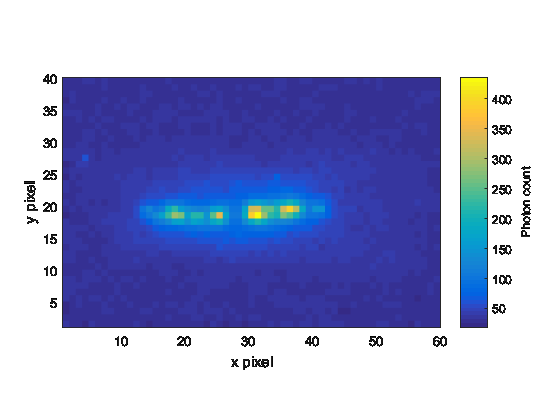}}
	\caption{
		Ion chain images from summing across all images from toggling the repump laser. 
		Each image is a separate experiment. 
		The first excitation laser ($\SI{554}{\nano\meter}$ laser) intensity is set to $\SI{0.935}{\watt\per\centi\meter\squared}$. 
		The total number of ions (bright and dark) are manually counted to be (a) 7, (b) 14, (c) 11, (d) 12, (e) 4, (f) 12, (g) 8, (h) 7, (i) 10, and (j) 8.}
	\label{fig:Ba137_Sum_images_36uW}
\end{figure*}
The experiments were also performed with the first excitation laser intensity turned up to $\SI{0.935}{\watt\per\centi\meter\squared}$. 
For this set of experiments, the number of repump laser toggling times is 10 for each ion chain instead of 100. 
At this first excitation laser intensity, the total number of ions trapped is counted to be 93, with 30 bright ions trapped in total (see Figure \ref{fig:Ba137_Sum_images_36uW}). 
This gives a selectivity of $32 \pm 5 \%$.

\section{$^{138}\mathrm{Ba}^+$ selectivity}
The $^{138}\mathrm{Ba}^+$ selectivity experiments are performed in the same way as described in Section \ref{sec:Ba-137 Selectivity}, but with the first excitation laser frequency set to the $^{138}\mathrm{Ba}^+$ isotope peak, and the cooling and repump laser frequencies set to the transition frequencies of $^{138}\mathrm{Ba}^+$ ions. 
53 sets of experiments were run for $^{138}\mathrm{Ba}^+$ selectivity, totalling 185 ions trapped, with 157 being bright. 
This gives a selectivity of $85 \pm 3 \%$.

\bibliography{bib}
\bibliographystyle{unsrt}
\end{document}